\documentclass[preprint,12pt]{elsarticle}

\usepackage{tabu}
\usepackage{array}  
\usepackage{amssymb}
\usepackage{placeins}
\usepackage{amsmath,epsfig}
\usepackage{setspace}
\usepackage{multirow}
\usepackage{float}
\usepackage{subfigure}
\usepackage{cellspace}
\usepackage{booktabs}
\usepackage{float}
\usepackage{makecell}
\usepackage{todonotes}
\usepackage{diagbox}
\usepackage{graphicx}
\usepackage{mathtools}
\usepackage{cases}
\usepackage{makecell}
\usepackage{geometry} 
\usepackage{lipsum}
\usepackage{breakurl}
\usepackage{url}

\journal{ISPRS}

\begin{document}

\begin{frontmatter}



\title{Towards Automatic SAR-Optical Stereogrammetry over Urban Areas using Very High Resolution Imagery}

\author{
Chunping Qiu\textsuperscript{a}, Michael Schmitt\textsuperscript{a}, Xiao Xiang Zhu\textsuperscript{a,b,}\corref{mycorrespondingauthor}}
		\cortext[mycorrespondingauthor]{Corresponding author}
\address{	\textsuperscript{a }Signal Processing in Earth Observation, Technical University of Munich (TUM), Munich, Germany\\
	\textsuperscript{b }Remote Sensing Technology Institute (IMF), German Aerospace Center (DLR), Wessling, Germany}

\newcommand\blfootnote[1]{% 
	\begingroup 
	\renewcommand\thefootnote{}\footnote{#1}% 
	\addtocounter{footnote}{-1}% 
	\endgroup 
}
\begin{abstract}
\renewcommand{\thefootnote}{}
\textit{This is the pre-print version, to read the final version please go to ISPRS Journal of Photogrammetry and Remote Sensing, Elsevier. (\url{https://doi.org/10.​1016/​j.​isprsjprs.​2017.​12.​006})}
		
In this paper we discuss the potential and challenges regarding SAR-optical stereogrammetry for urban areas, using very-high-resolution (VHR) remote sensing imagery. Since we do this mainly from a geometrical point of view, we first analyze the height reconstruction accuracy to be expected for different stereogrammetric configurations. Then, we propose a strategy for simultaneous tie point matching and 3D reconstruction, which exploits an epipolar-like search window constraint. To drive the matching and ensure some robustness, we combine different established hand-crafted similarity measures. For the experiments, we use real test data acquired by the Worldview-2, TerraSAR-X and MEMPHIS sensors. Our results show that SAR-optical stereogrammetry using VHR imagery is generally feasible with 3D positioning accuracies in the meter-domain, although the matching of these strongly hetereogeneous multi-sensor data remains very challenging.\blfootnote{ISPRS Journal of Photogrammetry and Remote Sensing, in press} 
\end{abstract}


\end{frontmatter}

\section{Introduction}\label{sec:Intro}

Currently, we are living in the ``golden era of Earth observation'',
characterized by an abundance of airborne and spaceborne
sensors providing a large variety of remote sensing data. In this
situation, every sensor type possesses different peculiarities,
designed for specific tasks. One prominent example is the
German interferometric SAR mission TanDEM-X, whose task is
the generation of a global Digital Elevation Model (DEM) \cite{Krieger2007}. In order to do so, for every region of interest highly
coherent InSAR image pairs acquired by the two satellites of this
mission are needed. The same holds for optical stereo sensors
such as RapidEye, which additionally require cloudless weather
and daylight during image acquisition \cite{Tyc2005}.
Eventually, this means that there is a huge amount of data in the
archives of which possibly a large potential remains unused,
because information currently can only be extracted within
those narrowly defined mission-specific configurations. If, e.g.,
a second coherent SAR acquisition is not (yet) available, or if one
image of an optical stereo pair is obstructed by severe cloud
coverage, the mission goal -- topography reconstruction -- can
currently not be fulfilled. The solution to this problem is the development of methods for flexible multi-sensor data fusion \cite{schmitt2016data}. This paper investigates the stereogrammetric fusion of SAR and optical imagery for a reconstruction of 3D information over urban areas less dependent on the type of the available remote sensing data.

The idea of SAR-optical stereogrammetry was first presented almost 30 years ago by Bloom et al. \cite{Bloom1988}, who investigated its general feasibility using low-resolution data provided by the SIR-B mission and the
Landsat-4/5 satellites, focusing on the analysis of rural areas. Further investigations were carried out by the group of Raggam et al. \cite{Raggam1990,Raggam1993,Raggam1994}, who combined low-resolution Seasat and SPOT/Landsat
images for rural DEM generation. Some time later, similar experiments using
ERS-2/Radarsat-1 and SPOT data were presented by \cite{Xing2008}. All these studies have shown errors in the dekameter-domain, thus seemingly prohibiting an application of SAR-optical
stereogrammetry to urban remote sensing. \textcolor{black}{While in \cite{wegnercombining} building height estimation by fusing a single-pass interferometric SAR image pair and one aerial orthophoto  has been shown to provide meter-accuracy, strict SAR-optical stereogrammetry was not studied by the authors.}  
Only recently, \cite{Zhang2015} showed that TerraSAR-X and GeoEye-1 images can be used to carry out stereogrammetric 3D reconstruction with an error
in the meter-domain -- although the study only used the manually measured corners of a simple-shaped building to proof the
concept. In contrast, based on the preliminary considerations sketched in \cite{Schmitt2016,Qiu2017}, our work intends to provide a first step towards a solution for the non-trival problem of
automatic stereo matching of VHR multi-sensor images of complex urban areas. In this context, it combines SAR-optical stereo intersection with a fully automatic selection of sparsely distributed tie points, i.e. it models both the matching and the reconstruction processes in a simultaneous manner. For this task, we first investigate the theoretically achievable accuracies of SAR-optical stereogrammetry, and how they depend on different intersection geometries. After that, we
propose an epipolar-like constraint for an enhancement of the difficult search for homologue tie points. To drive the matching and ensure some robustness, we combine different hand-crafted image descriptors and evaluate our findings on test data comprised of VHR remote sensing imagery acquired by the spaceborne sensors Worldview-2 and TerraSAR-X as well as the airborne system MEMPHIS.

\section{Principle of SAR-Optical Stereogrammetry}\label{sec:SAROPTprinciple}
\subsection{Geometric Interpretation}\label{sec:GeometricInterpretation}
\begin{figure}[htb]
\centering
\includegraphics[width=0.6\textwidth]{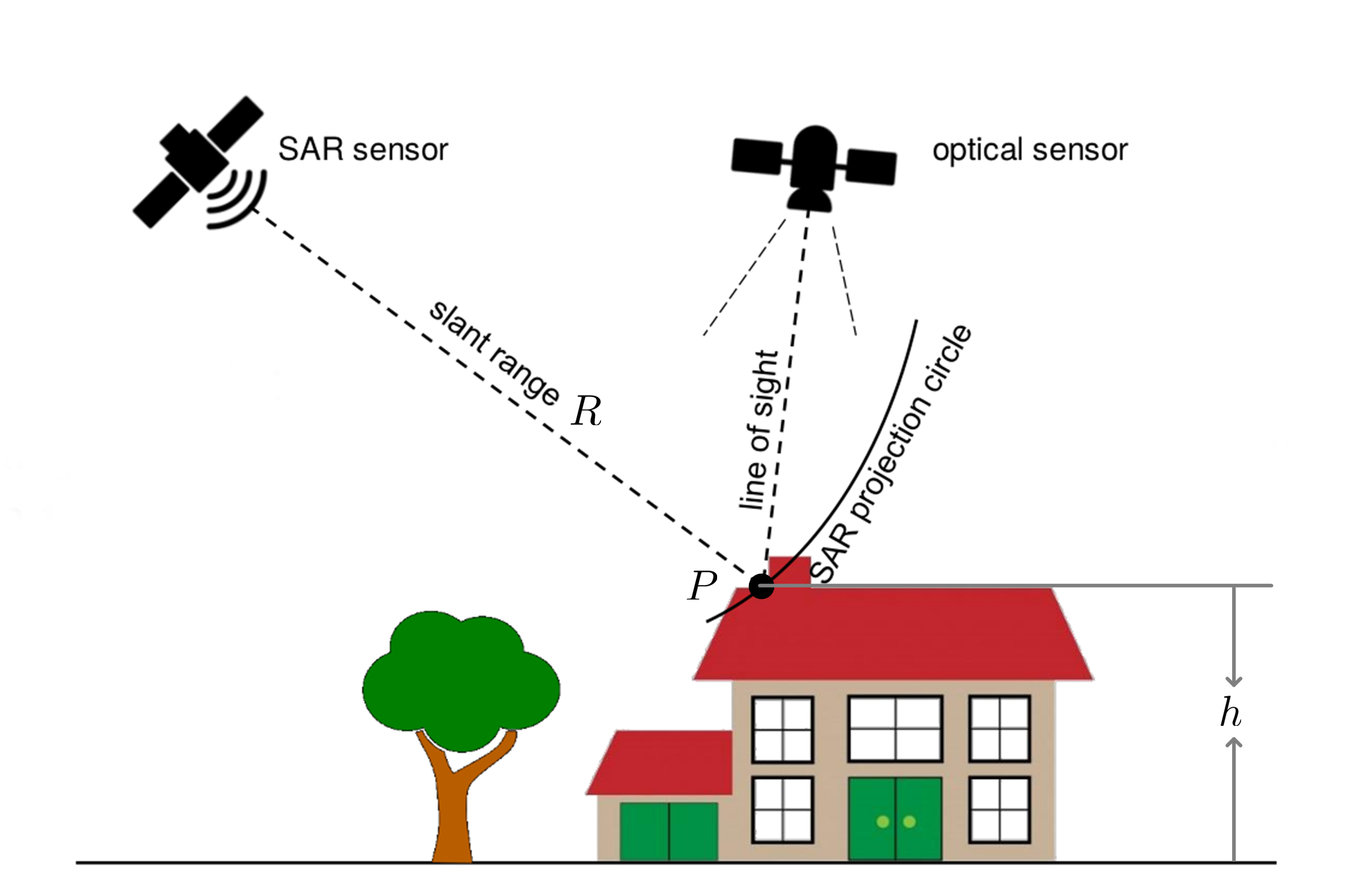}
\caption{The basic principle of SAR-optical stereogrammetry.}
\label{fig:principle}
\end{figure}
Figure~\ref{fig:principle} sketches the basic principle of SAR-optical stereogrammetry, which can be described as an intersection of the range-Doppler projection circle defined by the principal SAR measurements time $t$ and slant range $R$, and an optical projection ray defined by the optical image coordinates $x$ and $y$. Conceptually, this yields a set of four equations, namely the range-Doppler equations \cite{leberl1990radargrammetric}
\begin{equation}\label{eq:range}
R = \sqrt{\left(X_s(t) - X\right)^2 + \left(Y_s(t) - Y\right)^2 + \left(Z_s(t) - Z\right)^2}
\end{equation}
\begin{equation}\label{eq:Doppler}
V_x\left(X-X_s(t)\right) + V_y\left(Y-Y_s(t)\right) + V_z\left(Z-Z_s(t)\right)= 0
\end{equation}
of the zero-Doppler processed SAR data, and the central projection equations \cite{egels2003digital}
\begin{equation}\label{eq:optX}
x = x_0 + c\frac{r_{11}\left(X-X_o\right) + r_{21}\left(Y-Y_o\right) + r_{31}\left(Z-Z_o\right)}{r_{13}\left(X-X_o\right) + r_{23}\left(Y-Y_o\right) + r_{33}\left(Z-Z_o\right)} 
\end{equation}
\begin{equation}\label{eq:optY}
y = y_0 + c\frac{r_{12}\left(X-X_o\right) + r_{22}\left(Y-Y_o\right) + r_{32}\left(Z-Z_o\right)}{r_{13}\left(X-X_o\right) + r_{23}\left(Y-Y_o\right) + r_{33}\left(Z-Z_o\right)} 
\end{equation}
of the optical imagery. This overdetermined equation system can be solved for the unknown object coordinates $X$, $Y$ and $Z$, if the orientation parameters, i.e. sensor position $\mathbf{S}(t) = [X_s(t), Y_s(t), Z_s(t)]^T$ and instantaneous velocity $\mathbf{V}(t) = [V_x(t), V_y(t), V_z(t)]^T$ at zero-Doppler time $t$ of the SAR sensor, as well as projection center $\mathbf{P}_c = [X_o, Y_o, Z_o]$ of the optical sensor in addition to the elements of the rotation matrix $\mathbf{R} =  \begin{bmatrix}
r_{11} & r_{12} & r_{13} \\
r_{21} & r_{22} & r_{23} \\
r_{31} & r_{32} & r_{33} 
\end{bmatrix}$ depending on its orientation parameters $\phi, \omega, \kappa$, are known. 
A solution can be found, e.g. using least-squares estmation in the Gauss-Newton model.

\subsection{Theoretical Accuracy Analysis}\label{sec:analysis}
Obviously, the height accuracy that can be expected from SAR-optical stereogrammetry on the one hand depends on the accuracy of the observations, i.e. zero-Doppler time $t$ and slant range $R$ as measured by the SAR sensor and defined by the SAR image coordinates $(r,c)_s$, and the optical image coordinates $(r,c)_o$, which define the angular orientation of the optical projection ray. On the other hand, the accuracy also depends on the orbit and orientation parameters of the sensors. In this section, we derive some rules-of-thumb for the expectable accuracies depending on different acquisition configurations. Since the orbits of modern satellite missions are well-controlled, we focus on the modeling of the inaccuracies of the principal measurements, which define the intersection geometry given the sensor positions and orientations. 
\subsubsection{Opposite-Side Stereo}\label{sec:oppSide}
\begin{figure}[htb]
\centering
\includegraphics[width= 0.5\columnwidth]{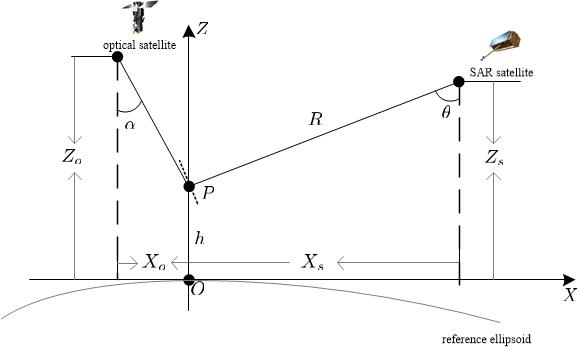}
\caption{SAR-optical intersection geometry for opposite-side stereo. The thin dotted line perpendicular to the look direction shows the SAR wave front, which can be considered as a tangent to the range-Doppler circle.} \label{fig:OppositeStereo}
\end{figure}
For sake of simplicity, Fig.~\ref{fig:OppositeStereo} sketches SAR-optical stereogrammetry as a trigonometric, in-plane intersection problem: Using $Z_s$ for the satellite height,%and $\theta$ for the actual radar viewing angle $\theta^{'}$,
 \textcolor{black}{the approximation error caused by ignoring the Earth curvature is $\left(\frac{Z_s-H_s}{H_s}\right)$, where $Z_s$ can be calculated given the radius of the reference ellipsoid $r$ and the radar viewing angle $\theta$. For  TerraSAR-X (TSX), $H_s = 515km$ and $\theta = {23}^{\circ}$, given $r = 6371km$, the approximation error} is within $0.8\%$. As explained in Section~\ref{sec:GeometricInterpretation}, the optical projection ray defined by the viewing angle $\alpha$ intersects with the range-Doppler circle of the SAR sensor, whose radius is defined by the range measurement $R$, which can also be represented by the radar viewing angle $\theta = \arccos\left(\frac{Z_s-h}{R}\right)$.

Thus, in a generalized manner, we can denote the height reconstruction process as a nonlinear function of $R$ and $\alpha$:
\begin{equation}
h = f(R,\alpha) 
\end{equation} 

If we then want to derive information about the height reconstruction accuracy $\sigma_h$, we can resort to variance-covariance propagation, which gives us
\begin{equation}\label{eq:propagation}
\sigma_{h}^2 = \left(\frac{\partial h}{\partial R}\right)^2\sigma_R^2 + \left(\frac{\partial h}{\partial \alpha}\right)^2\sigma_\alpha^2	
\end{equation} 
for uncorrelated measurements $R$ and $\alpha$. In order to get hold of the partial derivatives that are needed to calculate the height variance $\sigma_h^2$ using equation (\ref{eq:propagation}), we derive the following geometric relationships:

First of all, in the intersection plane ($Y=0$) sketched in Fig.~\ref{fig:OppositeStereo}, the sensor position can be described by
\begin{equation}\label{eq:oppSideXS}
X_s =   R sin(\theta),
\end{equation}
\begin{equation}\label{eq:oppSideZS}
Z_s =  H_s,
\end{equation}
while the position of the optical sensor can be calculated by
\begin{equation}\label{eq:oppSideXO}
X_o = -(H_o-h)tan(\alpha),
\end{equation}
\begin{equation}\label{eq:oppSideZO}
Z_o =  H_o.
\end{equation}
The projection ray corresponding to the optical measurement is defined by
\begin{equation}\label{eq:projectionRay}
 z = k(x-X_o) + Z_o.
\end{equation}
where $k$ is the slope of the line determined by the viewing angle of the optical sensor. For opposite-side stereo $k = -\cot(\alpha)$. Finally, the range equation for the SAR measurement is given by
\begin{equation}\label{eq:rangePlane}
(X_s-x)^{2} +  (Z_s-z)^{2} = R^{2}. 
\end{equation}

The combination of equations (\ref{eq:projectionRay}) and (\ref{eq:rangePlane}) yields a full description of the SAR-optical stereo intersection problem with two unknowns, namely $x$ and $z$. If we combine (\ref{eq:projectionRay}) and (\ref{eq:rangePlane}) into a single equation, i.e.
\begin{equation}
g =	 (X_s-(\dfrac{z- Z_o} {k} + X_o))^{2} +  (Z_s-z)^{2} - R^{2} = 0,
\end{equation} 
we can calculate the desired partial derivatives by 
\begin{equation}
\dfrac{\partial h}{\partial R}  = \dfrac{\partial z}{\partial R}   =   -\dfrac{ g_z} { g_R}  =  - \dfrac{ R k} {  (X_s-x) + k (Z_s - z) }   \end{equation} 
and  
\begin{equation}
\dfrac{\partial h}{\partial \alpha}  = \dfrac{\partial z}{\partial \alpha}   =   -\dfrac{ g_{\alpha}} { g_R} = - \dfrac{  (X_s-x)(Z_O-z)  } {  (X_s-x) + k (Z_s - z)  }  \dfrac{  1 } { k  }     \dfrac{\partial k}{\partial \alpha},
\end{equation} 
respectively.

\begin{figure} [H]  %[!tbh]   
 	\centering
 	\subfigure[]{\includegraphics[width= 6.5cm]{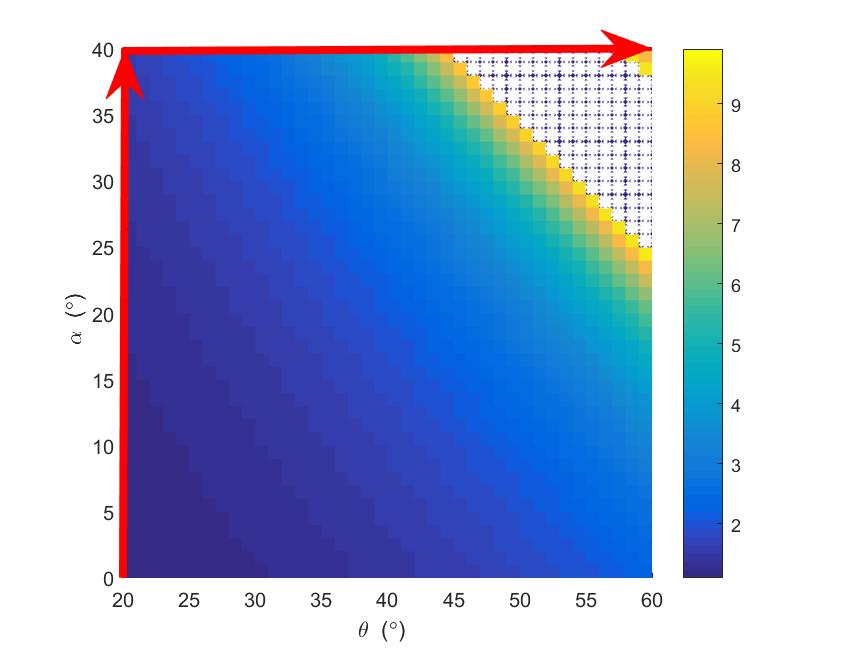}}
 	\subfigure[]{\includegraphics[width= 7.0cm]{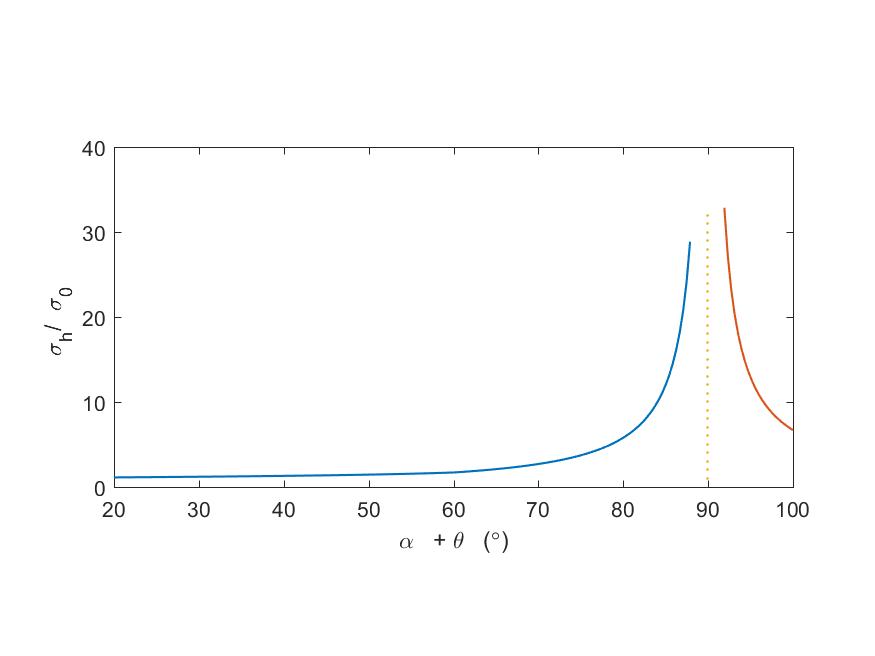}}
 	\caption{Normalized height accuracy $\frac{\sigma_h}{\sigma_0} $ for the opposite-side stereo case as a function of (a) $ \theta$ and $\alpha$, and (b) the intersection angle \textcolor{black}{of the sensor line of sights} $\left(\theta +　\alpha\right)$. The white stars in (a) mark the cases where $\frac{\sigma_h}{\sigma_0} > 10$. The \textcolor{black}{horizontal coordinate axis} $\theta +　\alpha$ in (b) corresponds to the sum of the axis values indicated by the red arrows in (a). } 
 	\label{fig:oppositeStereoAnalysis}
 \end{figure}

Given the standard deviations $\sigma_{R} = \sigma_{0}$ and $  \sigma_{\alpha} = 10^{-6}\sigma_{0}$,
the normalized height accuracy $\dfrac{\sigma_h}{\sigma_0} $ as a function of $\theta$, $\alpha$, and the intersection angle \textcolor{black}{of the sensor line of sights} $\left(\theta +　\alpha\right)$ can be seen in Figure~\ref{fig:oppositeStereoAnalysis}. It becomes obvious that \textcolor{black}{$\theta +　\alpha = 90^\circ$} does not allow proper stereo intersection anymore, which is explained by the geometric interpretation of the SAR-optical stereo configuration: \textcolor{black}{when $\theta +　\alpha = 90^\circ$}, the optical projection ray would act as a tangent to the range-Doppler circle defined by the SAR range measurement, thus leading to a glancing intersection. 
 
\subsubsection{Same-Side Stereo}\label{sec:sameSide}
\begin{figure}[htb]
\centering
\includegraphics[width= 0.5\columnwidth]{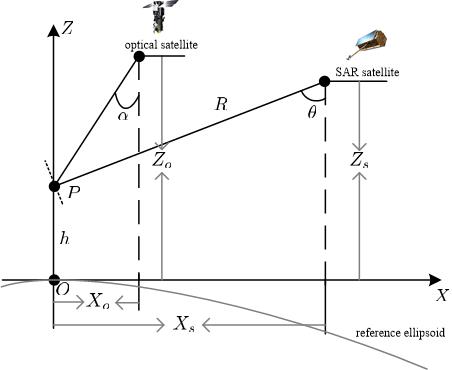}
\caption{SAR-optical intersection geometry for same-side stereo. The thin dotted line perpendicular to the look direction shows the SAR wave front, which can be considered as a tangent to the range-Doppler circle.}\label{fig:sameSideStereo}
\end{figure}
\begin{figure} [H]  %[!tbh]   
 	\centering
 	\subfigure[]{\includegraphics[width= 6.5cm]{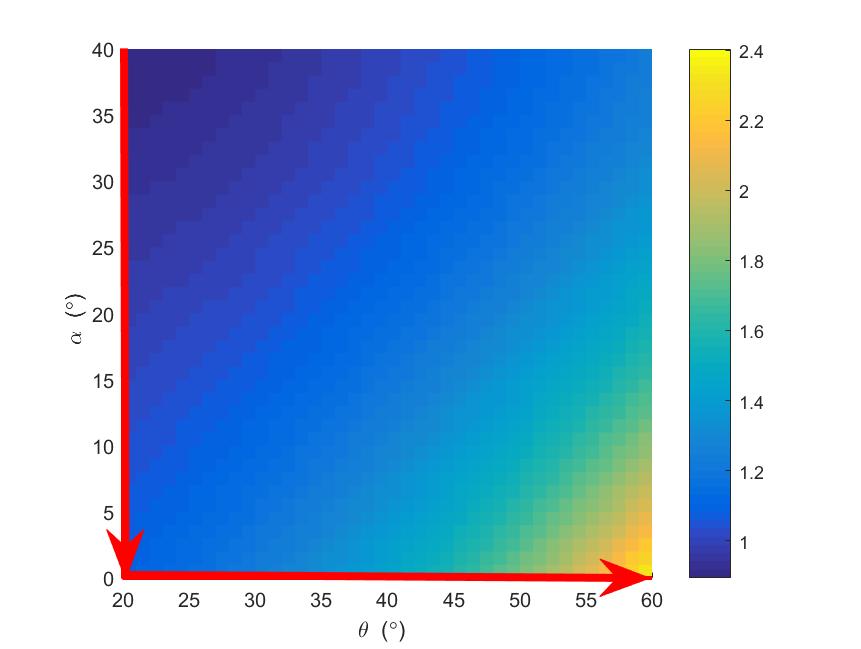}}
 	\subfigure[]{\includegraphics[width= 6.5cm]{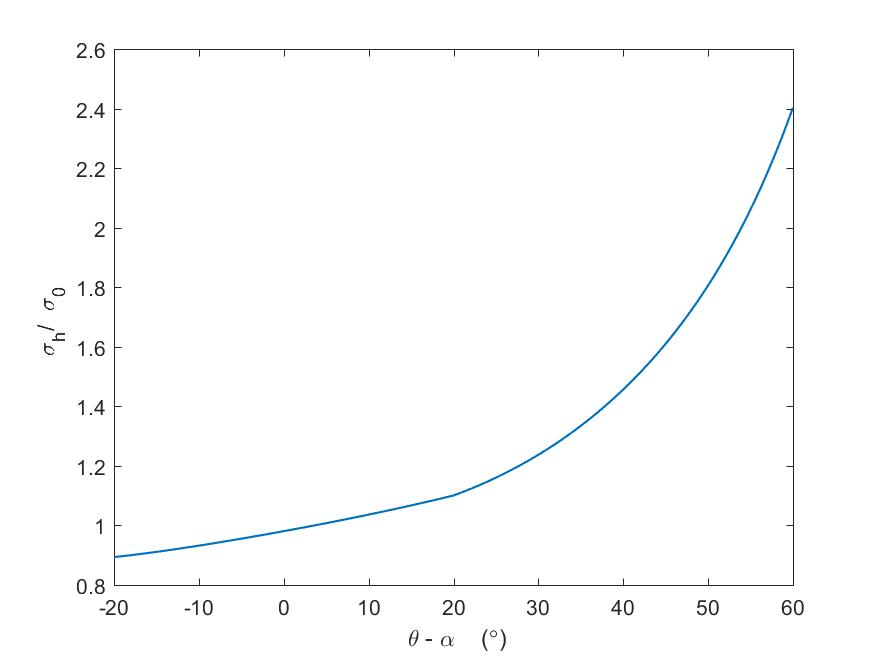}}
 	\caption{Normalized height accuracy $\dfrac{\sigma_h}{\sigma_0} $ for same-side stereo, as a function of $ \theta$, $\alpha$ (a), and the intersection angle \textcolor{black}{of the sensor line of sights} $\left(\theta -　\alpha\right)$ (b). The \textcolor{black}{horizontal coordinate axis} $\theta -　\alpha$ in (b) corresponds to the difference value of the axis values indicated by the red arrows in (a). A negative angle indicates the optical sensor in Figure~\ref{fig:sameSideStereo} is at the right side of the SAR sensor.}
 	\label{fig:sameStereoAnalysis}
 \end{figure}

 
In contrast to the opposite-side stereo case stands the same-side stereo case, which is sketched in Figure~\ref{fig:sameSideStereo}. Here, equation (\ref{eq:oppSideXO}) is replaced by 
\begin{equation}\label{eq:samSideXO}
X_o =  (H_o-h) \tan(\alpha)
\end{equation}
and $k = \cot(\alpha)$ in equation (\ref{eq:projectionRay}).
The normalized height accuracy $\frac{\sigma_h}{\sigma_0}$ for this case can be seen in Figure~\ref{fig:sameStereoAnalysis}. Here, it becomes obvious that the intersection geometry is best for small intersection angles \textcolor{black}{of sensor line of sights}, which will lead to a perfect, perpendicular intersection of optical projection ray and range-Doppler circle.
   
 Besides providing us some rules-of-thumb for the expectable SAR-optical sterogrammetry accuracies with respect to different acquisition conﬁgurations, the considerations in this section also give us an indication of the potential SAR-optical sterogrammetry accuracy: an accuracy on the same order of the slant range $R$ accuracy is possible. For TerraSAR-X and TanDEM-X, this accuracy can theoretically be at the cm level \cite{zhu2016geodetic}. For medium-resolution sensors, such as Sentinel-1, the positioning capability is less precise.

\FloatBarrier
\section{Stereogrammetry by Simultaneous Tie Point Matching and 3D Reconstruction}\label{sec:Matching}
\begin{figure}[!htb]
\centering
\includegraphics[width=0.7\textwidth]{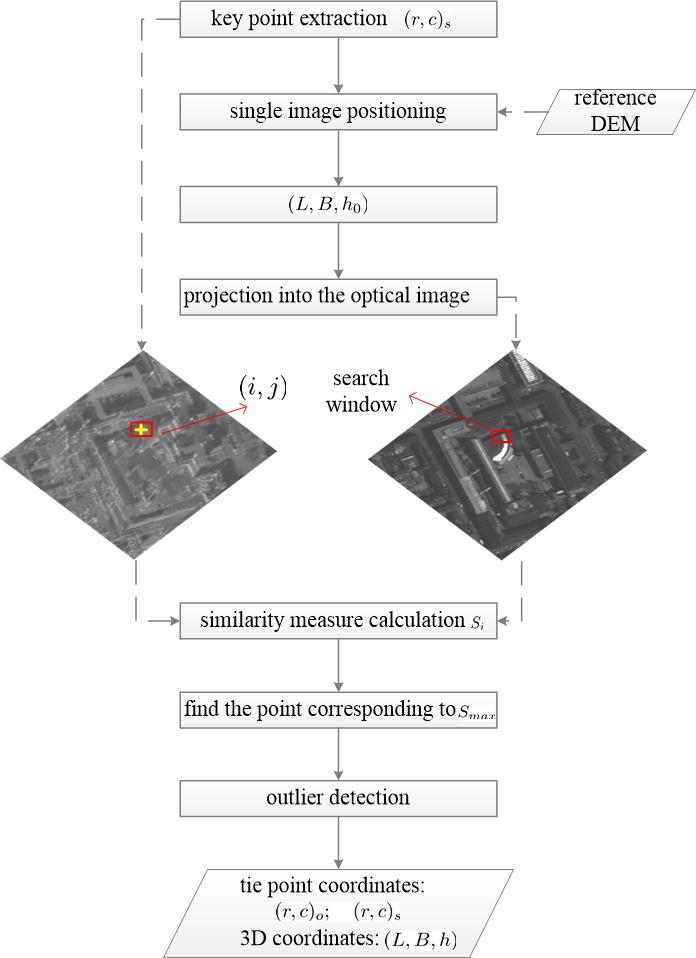}
\caption{Procedure of the proposed tie point matching strategy.}
\label{fig:flowchart}
\end{figure}

Since the correct matching of homologue points is a very difficult task for VHR multi-sensor remote sensing images, we propose a framework that carries out both the tie point matching and the 3D-reconstruction steps in a joint, iterative manner. The framework's major strength is that it reduces the search space dramatically by incorporating light prior knowledge about the scene of interest and a constraint inspired by epipolar geometry. The algorithm for tie point selection is sketched in Fig.~\ref{fig:flowchart}. 
It consists of the following steps:
\begin{enumerate}	
\item{First, a set of keypoints $(r, c)_{s}$ is detected in the SAR image using a block-based Harris corner detector \cite{Ye2017}. Beginning with key point detection in the SAR image instead of the optical image is meant to avoid situations in which a ground point does not exist in the SAR image at all, because of radar shadow or layover.}
\item{The detected keypoints are then projected onto a coarse elevation model (e.g. the SRTM DEM) using classical single-image positioning. This yields an approximate longitude, latitude and height coordinate $(L, B,h_0)$ for every key point. For cities residing on rather flat topography and containing mostly smaller buildings, a simple plane located at the average scene height could be used as coarse elevation model.}
\item{ For every keypoint, $(L, B, h_0)$ is then projected into the optical image, yielding the corresponding image coordinates $(r, c)_{o}$. 
Around this image point, a search window is established, whose details are described in Section \ref{sec:SearchWindow}.}
\item{Different similarity measures $S_i$ are used to compare all pixels in this search window to the original SAR image key point. The optical image point $(r, c)^{m}_{o}$ corresponding to the maximum similarity $S_m$ is considered as the potential homologue point of the SAR key point.}
\item{
In order to exclude possible mis-matched tie points, matching reliability is enhanced using a joint exploitation of several similarity measures, which only keeps tie points if they are supported by several different measures.    

}
\end{enumerate}

The final output of this procedure is the image coordinates $(r, c)_{o}$ and $(r, c)_{s}$ of the tie points as well as the 3D coordinates $(L, B, h)$ of the corresponding object point, i.e. both matching and 3D-reconstruction are solved in a joint, simultaneous manner. The individual processing steps are described in more detail in the following sections.

\subsection{Restricting the Search Space}\label{sec:SearchWindow}
 
The epipolar-line constraint is well-known in classical photogrammetry \cite{Zhang1995} and serves as an efficient search strategy in stereo image matching as it reduces the search space for homologue points from two dimensions to only one dimension. Unfortunately, a rigorous epipolar-line does not exist for SAR stereo pairs \cite{Gutjahr2014}, and even less so for SAR-optical stereo pairs. Therefore, in this paper, we propose a similar search strategy specifically designed for SAR-optical image matching. We call the result \textit{Imaging-Model-based-Line-Shape} (IMBLS) search window. As shown in the following section, this strategy can be derived by theoretical considerations of the imaging geometry.
\subsubsection{Derivation of a Search Line Constraint}\label{sec:window1}
\noindent 
For simplicity and generality, the imaging models of SAR and optical images, respectively, can be expressed using the general non-linear functionals 
\begin{equation}\label{equ:LBH2rcSar1}
r_s = f_{1}(\mathbf{p_s}; L,B,h)
\end{equation}
\begin{equation}\label{equ:LBH2rcSar2}
c_s = f_{2}(\mathbf{p_s}; L,B,h)
\end{equation}
and
\begin{equation}\label{equ:LBH2rcOpt1}
r_o = f_{3}(\mathbf{p_o}; L,B,h)
\end{equation}
\begin{equation}\label{equ:LBH2rcOpt2}
c_o = f_{4}(\mathbf{p_o}; L,B,h),
\end{equation}

where $(r_s,c_s)$ are the tie point coordinates in the SAR image, $\mathbf{p_s}$ are the SAR orientation parameters and consist of the instantaneous sensor position and velocity at time $t$, $\mathbf{p_s} = [\mathbf{S}(t_i), \mathbf{V}(t_i) ]$ \cite{leberl1990radargrammetric}; $(r_o,c_o)$ are the tie point coordinates in the optical image, and $\mathbf{p_o}$ are the optical orientation parameters that consist of sensor position and orientation, $\mathbf{p_o} = [X_o, Y_o, Z_o, \phi, \omega, \kappa]$ \cite{xing2013bundle}. $(L,B,h)$ are the 3D coordinates of the ground point corresponding to this pair of tie points. 
The derivation starts from the assumption that a keypoint $(r,c)_s$ has been detected in the SAR image for which the corresponding point $(r,c)_o$  in the optical image needs to be found in order to perform stereogrammetric forward intersection. Taking both steps, i.e. image matching and 3D reconstruction into consideration simultaneously, the problem can be formulated as a system of four equations with five unknowns: the object coordinates $(L,B,h)$ and the optical tie point image coordinates $(r,c)_o$. Although a solution is not possible in a straight-forward manner, a constraint between the unknowns can be constructed as follows. First, (\ref{equ:LBH2rcSar1}) and (\ref{equ:LBH2rcSar2}) can be rewritten as
\begin{equation}\label{equ:rch2LBSar1}
L = g_{1}( \mathbf{p_s}; h,r_s,c_s)
\end{equation}
\begin{equation}\label{equ:rch2LBSar2}
B = g_{2}( \mathbf{p_s}; h,r_s,c_s),
\end{equation}

(\ref{equ:LBH2rcOpt1}) and (\ref{equ:LBH2rcOpt2}) can be rewritten as
\begin{equation}\label{equ:rch2LBOpt1}
h = g_{3}(\mathbf{p_o}; L, B, r_o)
\end{equation}
\begin{equation}\label{equ:rch2LBOpt2}
h = g_{4}(\mathbf{p_o}; L, B, c_o).
\end{equation}

where $g_1$ and $ g_2 $ describe the mapping of the keypoint coordinates $(r,c)_s$ to a spatial position  $(L,B)$ for a given height $h$, and $g_3$ and $g_4$ in analogy are nonlinear functions to derive the height $h$ if the spatial position $(L,B)$ and the optical coordinates $(r,c)_o$ are known.

Replacing the ground coordinates $L$ and $B$ in (\ref{equ:rch2LBOpt1}) and (\ref{equ:rch2LBOpt2}) by (\ref{equ:rch2LBSar1}) and (\ref{equ:rch2LBSar2}) and combining (\ref{equ:rch2LBOpt1}) and (\ref{equ:rch2LBOpt2}), we get the equality
\begin{equation}\label{equ:hEqualh}
g_{3}(\mathbf{p_o}; g_{1}, g_{2} ,r_o) =  g_{4}(\mathbf{p_o}; g_{1}, g_{2}, c_o).
\end{equation}
Finally, from (\ref{equ:hEqualh}) we can acquire a constraint $G$ on the row coordinate $r_o$ corresponding to the column coordinate $c_o$ of the homologue point in the optical image by
\begin{equation}%\label{equ:6}
r_o = G(\mathbf{p_o},\mathbf{p_s},r_s,c_s, L,B,h; c_o).
\end{equation}
Even though we are not deriving a concise analytic expression for the constraint $G$, it provides a search line analogue to the epipolar constraint, i.e. a line along which the optical tie point to a given key point in the SAR image must reside. A similar constraint can be derived when starting from key points detected in the optical image. 

\subsubsection{Construction of the IMBLS Search Window}\label{sec:window2}
\begin{figure}[htb]
\centering
\subfigure[An exemplary key point detected in the SAR image.]{\includegraphics[width=0.28\textwidth]{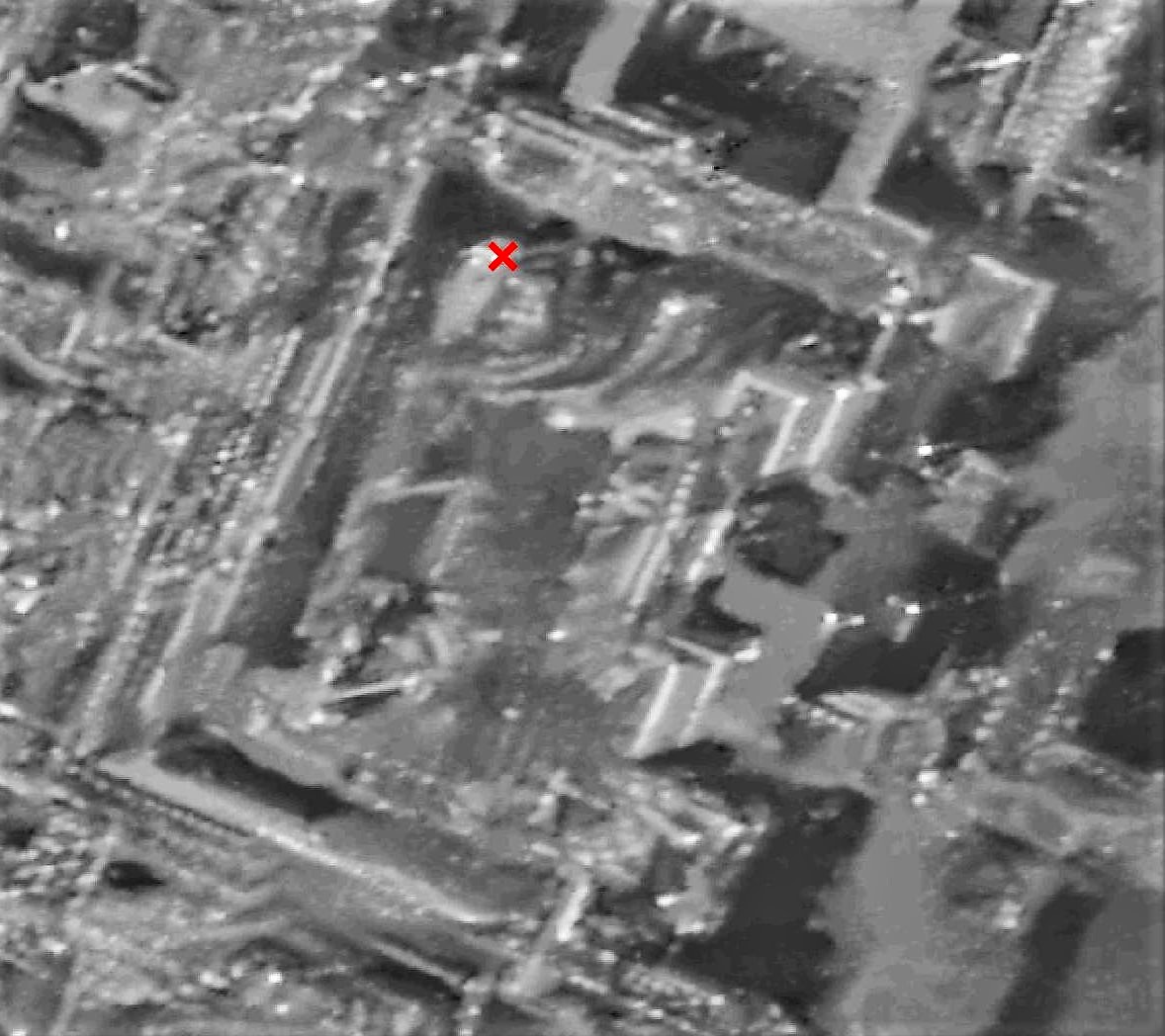}\label{fig:linshape1Keypoint}}
\subfigure[Sketch of the procedure to calculate the optical image coordinates for every height $h_k$.]{
\includegraphics[width=0.3\textwidth]{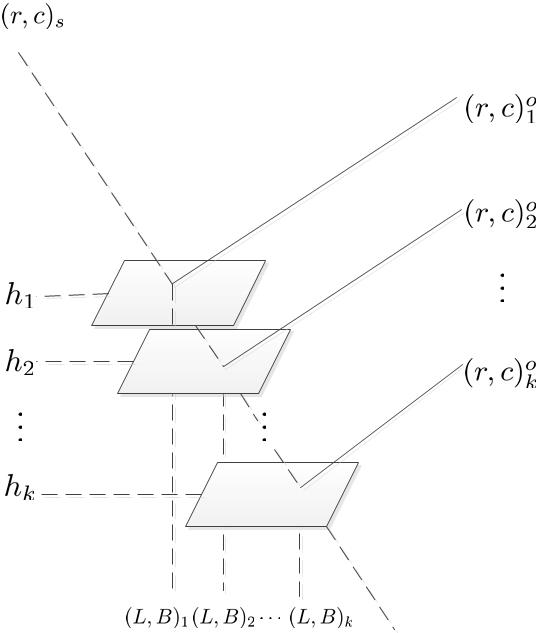}\label{fig:linshape2Procedure}
}
\subfigure[The projection line described by $(r,c)^{o}_{k}$ in the optical image.]{		\includegraphics[width=0.28\textwidth]{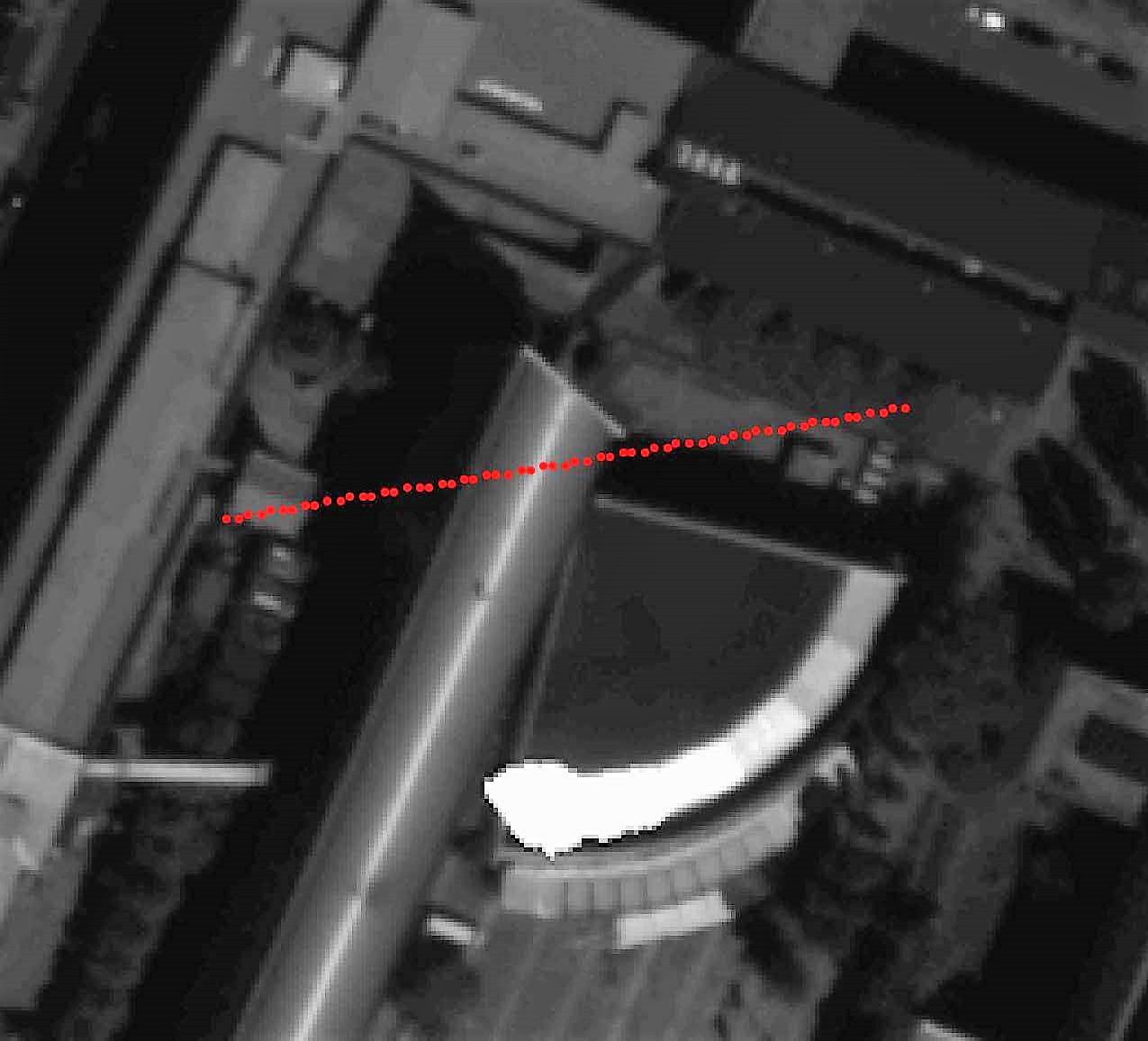}
\label{fig:linshape3Line}
}
\caption{Exemplary illustration of the procedure for construction of the IMBLS search window.}
%\label{fig:lin1}
\end{figure}

\begin{figure}[htb]
	\centering
	\includegraphics[width=0.45\textwidth]{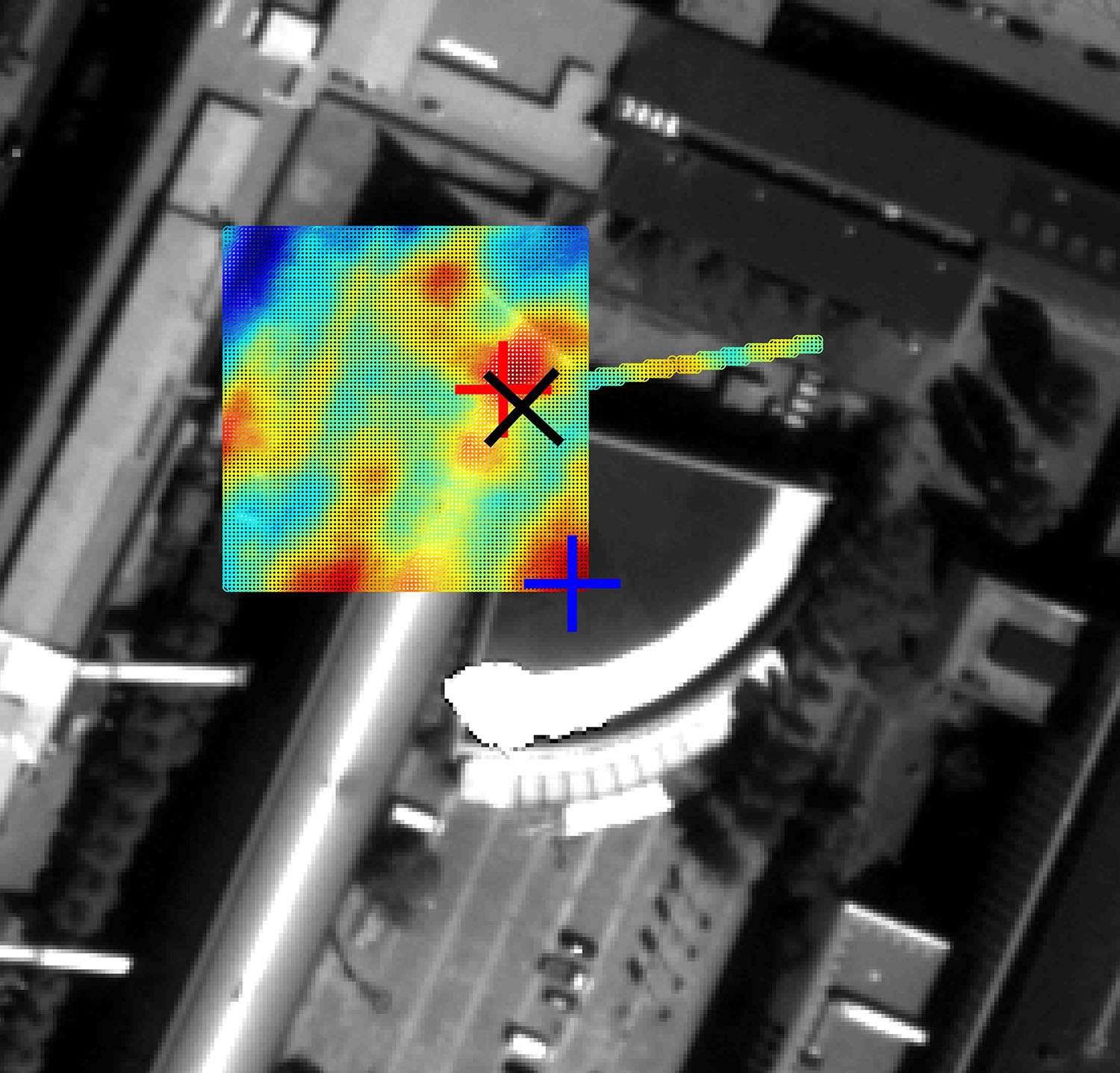}
	\caption{Similarity values calculated for the classical rectangular window and the IMBLS window, colorized by the similarity measure. The red and blue "+" marks indicate the pixels with the maximum similarity within the classical rectangular and the IMBLS window, respectively. The black "x" indicates the correct homologue point.}
	\label{fig:linshapeRegualarWindowEffect}
 \end{figure}

Exploiting the constraint derived in the previous section, a line-shaped search window can be constructed. As Fig.~\ref{fig:linshape1Keypoint}  shows, the procedure starts from an arbitrary key point $(r,c)_s$ detected in the SAR image. Using (\ref{equ:rch2LBSar1}) and (\ref{equ:rch2LBSar2}), and a suitable height search space $h_k \in [h_{min},h_{max}]$ , a set of potential spatial ground coordinates $(L,B)_k $ can be calculated for every  $h_k $, and the corresponding optical image coordinates  $(r,c)_k^o $ can be calculated using (\ref{equ:LBH2rcOpt1}) and (\ref{equ:LBH2rcOpt2}). This basic principle is illustrated in Fig.\ref{fig:linshape2Procedure}, while the resulting search line described by $(r,c)_k^o $ can be seen in Fig.\ref{fig:linshape3Line}.
As discussed in Section~\ref{sec:SearchWindow}, the tie point corresponding to the original SAR key point theoretically needs to be located along this line. However, in order to deal with possible inaccuracies in the imaging parameters $\mathbf{p_o}$ and $\mathbf{p_s}$, a pre-defined buffer width needs to be added around the line. An exemplary IMBLS search window and a regular square search window around the candidate homologue point in the optical image are compared in Fig.~\ref{fig:linshapeRegualarWindowEffect}, which also illustrates how quickly image matching can fail if the search window is too large and thus allows ambiguous results. The proposed IMBLS window restricts the search space significantly, thus making the image matching both more reliable and faster.

Besides a reduced search space, an important advantage of the IMBLS window is that every candidate point located in the window corresponds to a set of 3D coordinates ($L,B,h$), which means that 3D-reconstruction of object coordinates is solved simultaneous to tie point matching.

\subsection{Similarity Measures for SAR-Optical Imagery Matching}\label{sec:SimilarityMeasures}
In the context of this paper, we follow the generic approach of Inglada \& Giros \cite{Inglada2004}, who define the similarity measure between two images $I$ and $J$ as a strictly positive scalar function
\begin{equation}%\label{equ:1}
S_c(I,J) = f(I,J;c),
\end{equation}
where $c$ is a to-be-defined similarity criterion. $S_c$ has the maximum when the two images are identical according to the similarity criterion. In the framework of this paper, we extend this definition by allowing negative values so that similarity measure such as the correlation coefficient whose value range by definition is $[-1;+1]$, can be considered as a similarity measure as well.

\subsubsection{Signal-based Similarity Measures}\label{sec:sigBased}
 Signal-based similarity measures 

are calculated based on the original or pre-processed signals, i.e. gray values of pixels in the image processing case.
In this paper, we investigate two widely used measures:
\begin {itemize}
\item
\textit{Normalized Cross-Correlation (NCC)}\\
The normalized cross-correlation coefficient
\begin{equation}
\rho\left(x,y\right) = \frac{1}{N-1}\sum_{x,y}{\frac{\left(I\left(x,y\right)-\bar{I}\right)\left(J\left(x,y\right)-\bar{J}\right)}{\sigma_I\sigma_J}}
\end{equation}
%\todo[inline]{Please check if formula is correct. I think so.}
correlates two image patches $I$ and $J$, where $N$ is the number of the pixels in the image patch, while implicitly normalizing them to reduce the effects of changing image brightness.
\item
\textit{Mutual Information (MI)}\\
Mutual information is defined as the function 
of the joint entropy $H(I,J)$ and the marginal entropies $H(I), H(J)$ of two images $I$, $J$. We employ its normalized version in this paper \cite{studholme1999overlap}.
\end{itemize}

\subsubsection{Descriptor-based Similarity Measures}\label{sec:descBased}
Image descriptors are a well-established means to describe images on a global as well as a local scale. In the context of image matching, usually local descriptors are extracted around previously detected key points. Subsequently, the resulting feature vectors are compared using a suitable distance metric. In the scope of this paper, we resort to the negative $L_2$-norm as similarity metric. We chose the following descriptors in this investigation:
\begin{itemize}
\item 
\textit{Histogram of Oriented Gradients (HOG)}\\
The HOG descriptor was first proposed in 1986 \cite{mcconnell1986method} in the context of object detection. Its principle is to count occurrences of gradient orientation on a dense grid of uniformly spaced image cells, using overlapping local contrast normalization for improved accuracy. 
\item
\textit{Scale-Invariant Feature Transform (SIFT)}\\
SIFT \cite{lowe2004distinctive} is the most prominent example of a local feature descriptor that has found wide application in the fields of computer vision and optical image analysis for more than a decade. The SIFT feature vector usually contains 128 elements depicting the normalized values of previously computed orientation histograms -- an analogy to HOG. In its original implementation, SIFT combines both feature point detection and descriptor extraction, so that the feature vector corresponds to a specific scale and orientation assigned to the detected key point. In this paper, we calculate the descriptor for a fixed scale and orientation of 10 and zero respectively.
 \item
\textit{Histogram of Orientated Phase Congruency (HOPC)}\\
HOPC \cite{Ye2017} is a relatively new local image descriptor that is also based on the analysis of oriented histograms, although the descriptor vector here is calculated from phase congruency \cite{kovesi2000phase} instead of gradient information. \textcolor{black}{HOPC is built by extending the phase congruency model with illumination and contrast invariance, in order to capture geometric structure or shape features of images. \cite{Ye2017} shows that HOPC is able to represent geometric structural similarities between multi-modal remote sensing images and is robust against significant non-linear radiometric changes.} That makes it supposedly well-suited to the case of multi-sensor image analysis. 
\end{itemize}

\subsubsection{Outlier Removal by Joint Exploitation of Several Similarity Measures}\label{sec:outlier} 
%\todo[inline]{\textcolor{green}{how to cite IGARSS paper?}}
\begin{figure}[htb]%[!tbh]
 	\begin{center}
 		\includegraphics[width= 1 \columnwidth]{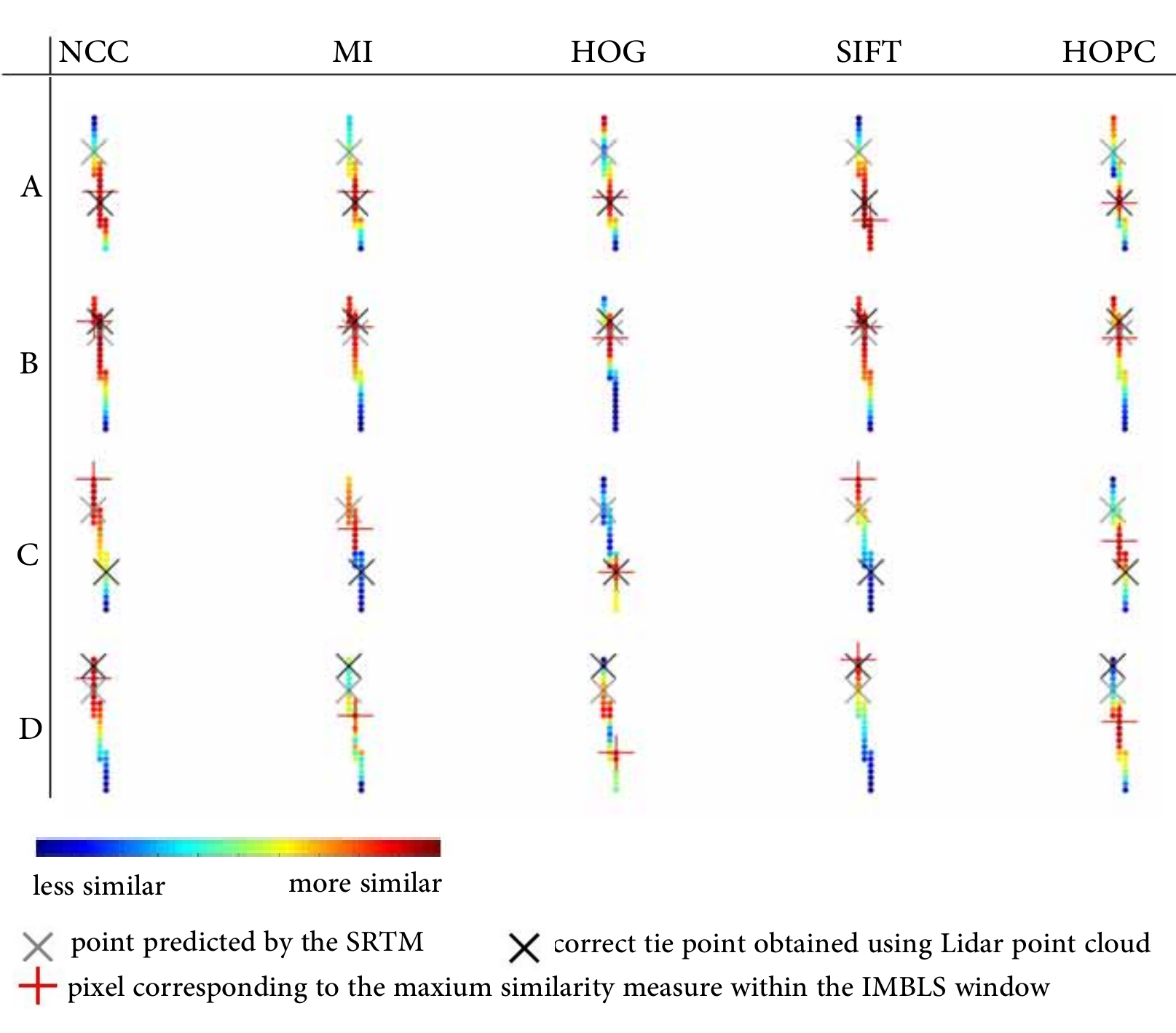}
 		\caption{Five similarity measures of four exemplary key points in the IMLBS search windows. \textcolor{black}{The points are colorized according to the calculated similarity measure.}}
 		 \label{fig:similarityMesuresEffect2Keypoints}
 	\end{center}
 \end{figure} 
Figure~\ref{fig:similarityMesuresEffect2Keypoints} shows a comparison of the similarity measures described in the previous subsections for four randomly chosen, exemplary key points A, B, C, and D. Although each measure shows some successful cases, no measure provides a perfect solution to any matching situation.
%, as also indicated in the previous study \cite{}
\textcolor{black}{As a first shot towards improved SAR-optical image matching, we} thus combine the individual similarity measures to enhance the robustness of the subsequent 3D reconstruction \textcolor{black}{by exploiting the individual capabilities to model similarities of the two signal-based similarity measures and the three different HOG-derivatives.}. 
For this combination, we simply calculate 
    \begin{equation}
D_{outlier} = max{(r_i)} - min{(r_i)} + max{(c_i)} - min{(c_i)},
\end{equation}%\mathbf
where $i\in {[NCC, MI, HOG, SIFT, HOPC ]}$, and $(r_i, c_i)$ are the coordinates of the pixels deemed most similar by similarity measure $i$. The smaller the value of $D_{outlier}$, i.e. the closer the individual similarity maxima are located, the more reliable is the matching result, as it indicates that the individual similarity measures are in agreement regarding the point under investigation. Staying with the example of Fig.~\ref{fig:similarityMesuresEffect2Keypoints} and given a threshold of \textcolor{black}{5} pixels,  \textcolor{black}{for example}, only the points A and B would finally be kept, while the other ones would be discarded.

\FloatBarrier
\section{Experiments and Results}\label{sec:Experiments}
\subsection{Test Data} \label{sec:data}

\begin{figure}[H]
\centering
\includegraphics[width= 0.5\columnwidth]{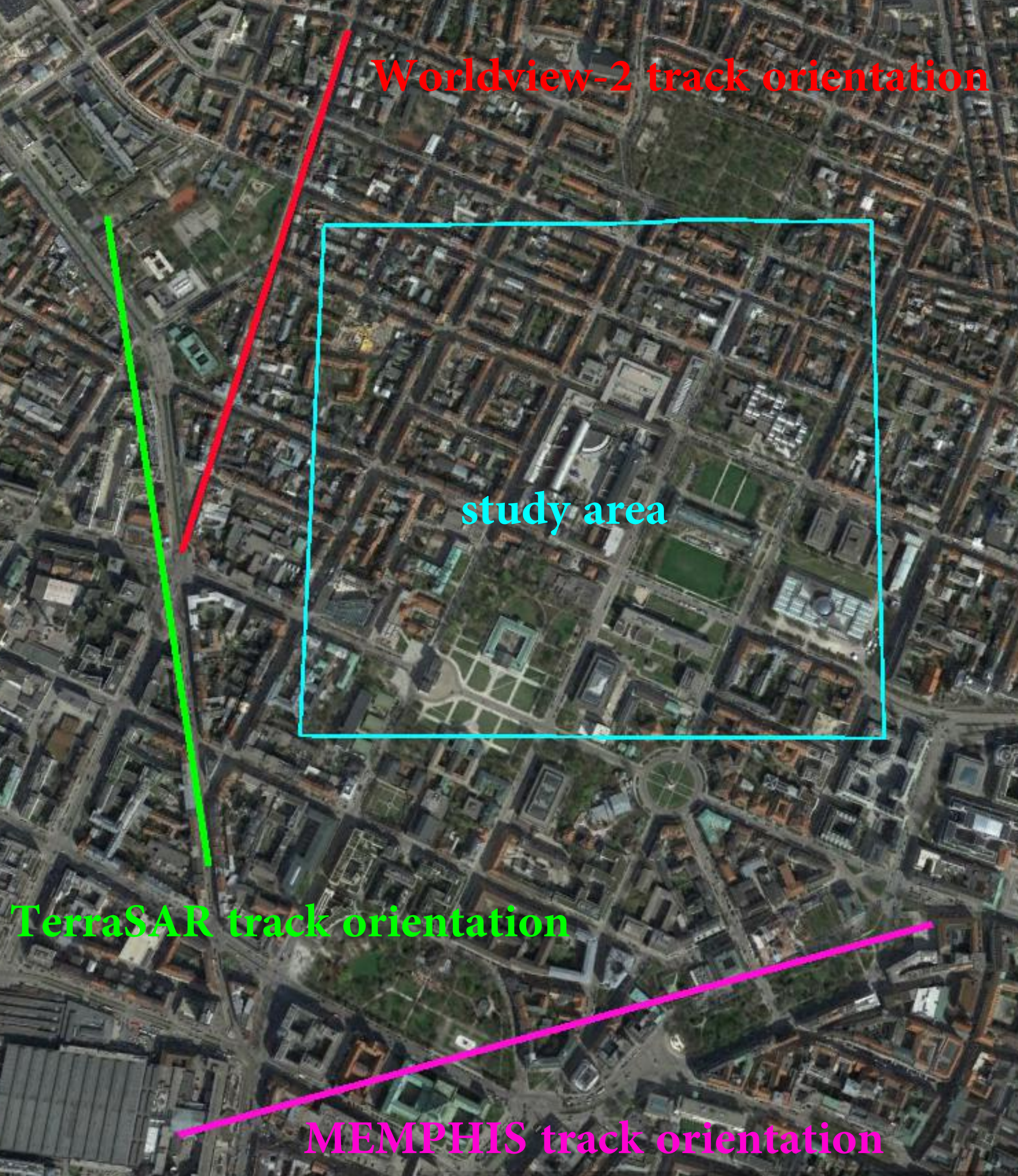}
\caption{The track relation of the three sensors (in Google Earth) over Munich.} \label{fig:areaMunich}
\end{figure}

\textcolor{black}{We present experiments for two areas of interest in Germany: Munich and Berlin. For them, several different VHR remote sensing images were acquired.} Details of the Munich dataset, which consists of images acquired by the spaceborne optical WorldView-2 (WV2) sensor, as well as the spaceborne SAR sensor TerraSAR-X (TSX) and the airborne SAR sensor MEMPHIS, are summarized in Tab.~\ref{tab:imgparaMunich}. The relation of the track orientations of the three acquisitions is shown in Fig.~\ref{fig:areaMunich}. 
\begin{table*}[tbh]
	\newcommand{\tabincell}[2]{\begin{tabular}{@{}#1@{}}#2\end{tabular}}
		\centering
	\caption{Parameters of the test images over Munich, Germany}
	\begin{tabular}{cccc}
		\toprule
        
scene&  \multicolumn{3}{c}{Munich}  \\
\cmidrule{2-4}  
image& {WorldView-2}& {TerraSAR-X} & {MEMPHIS} \\
\midrule  
acquisition date&{2015.07.22} &  {2015.03.04}  &  {2013.06.11 } \\ 
carrier platform & spaceborne & spaceborne & airborne\\
acquisition mode & panchromatic & staring spotlight & stripmap\\
mean off-nadir angle  &  ${13.7}^{\circ}$  & ${23}^{\circ}$   & ${55}^{\circ}$  \\
 	\multirow{2}[2]{*}{\tabincell{c}{pixel\\ spacing}  }  
   & {$0.49m$}& { $16.75cm (a)$} & {$ 5.21cm(a)$ }\\ 
   & {$0.49m$}& { $58.85cm (r) $} & {$ 16.65cm (r)$ }   \\
   band & visible & X-band & Ka-band\\
orbit height &  {$770km$}  &  {$515km$} &  {$760m$}  \\
    %  preprocessed pixel spacing& {$0.49m$}& { $\sim0.49m $}  &  {$ \sim0.49m$ }  \\
       applied rotation&-& $  {11}^{\circ}$ & $ {120}^{\circ}$ \\
		\bottomrule
	\end{tabular}%
	\label{tab:imgparaMunich}%
\end{table*}%

\begin{figure}[H]
\centering
\includegraphics[width= 0.4\columnwidth]{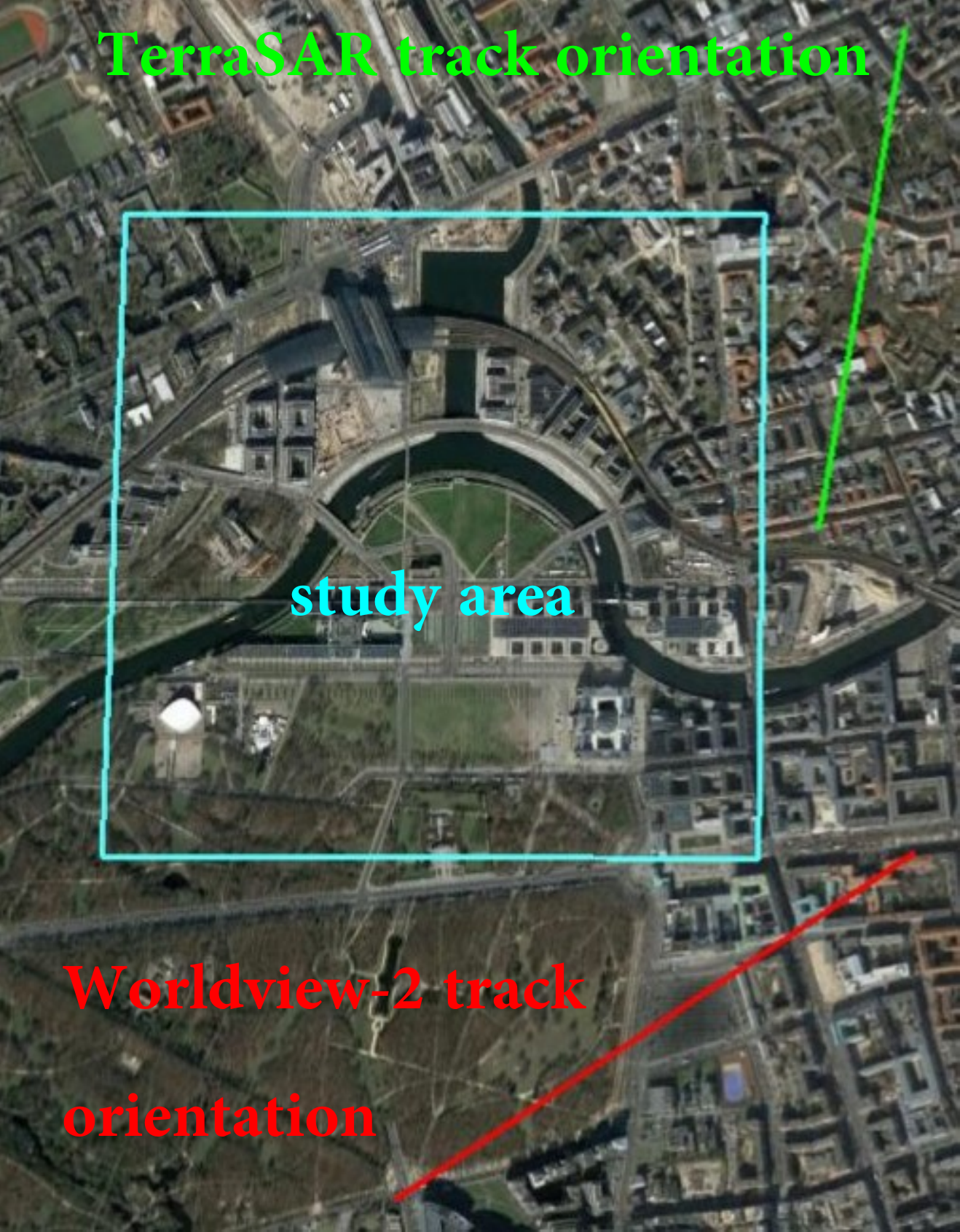}
\caption{The track relation of the two sensors (in Google Earth) over Berlin.} \label{fig:areaBerlin}
\end{figure}

\textcolor{black}{In addition, two VHR remote sensing images acquired over the city of Berlin, Germany, were used, whose details are summarized in Tab.~\ref{tab:imgparaBerlin}. Again, the optical image was acquired by the spaceborne optical WorldView-2 (WV2) sensor, and the SAR image by the spaceborne SAR sensor TerraSAR-X (TSX). The relation of the track orientations of the two acquisitions is shown in Fig.~\ref{fig:areaBerlin}.}

\begin{table*}[tbh]
	\newcommand{\tabincell}[2]{\begin{tabular}{@{}#1@{}}#2\end{tabular}}
		\centering
	\caption{Parameters of the test images over Berlin, Germany}
	\begin{tabular}{ccc}
		\toprule
        	scene&  \multicolumn{2}{c}{Berlin} \\
             %   \cmidrule{5-6}  
	image& {WorldView-2}& {TerraSAR-X} \\
    \midrule
    
acquisition date  &  {2013.05.05}  &  {2016.04.11}\\ 
carrier platform& spaceborne & spaceborne\\
acquisition mode & panchromatic & staring spotlight\\
mean off-nadir angle   & ${29}^{\circ}$   & ${36}^{\circ}$  \\
 	\multirow{2}[2]{*}{\tabincell{c}{pixel\\ spacing}  } 

  & { $0.65m (row)$} & {$16.69cm(a)$ }\\ 
  & { $0.55m (col)$} & {$ 58.85cm(r)$ } \\
   
   band & visible & X-band\\
orbit height &    {$770km$}  &  {$515km$}\\
    %  preprocessed pixel spacing& {$0.49m$}& { $\sim0.49m $}  &  {$ \sim0.49m$ }  \\
       applied rotation&- & $ {11}^{\circ}$ \\
        
		\bottomrule
	\end{tabular}%
	\label{tab:imgparaBerlin}%
\end{table*}%

\begin{figure}[htb]  
 	\centering
 	\subfigure[\textcolor{black}{MEMPHIS patch}]{\includegraphics[height=4.3cm]{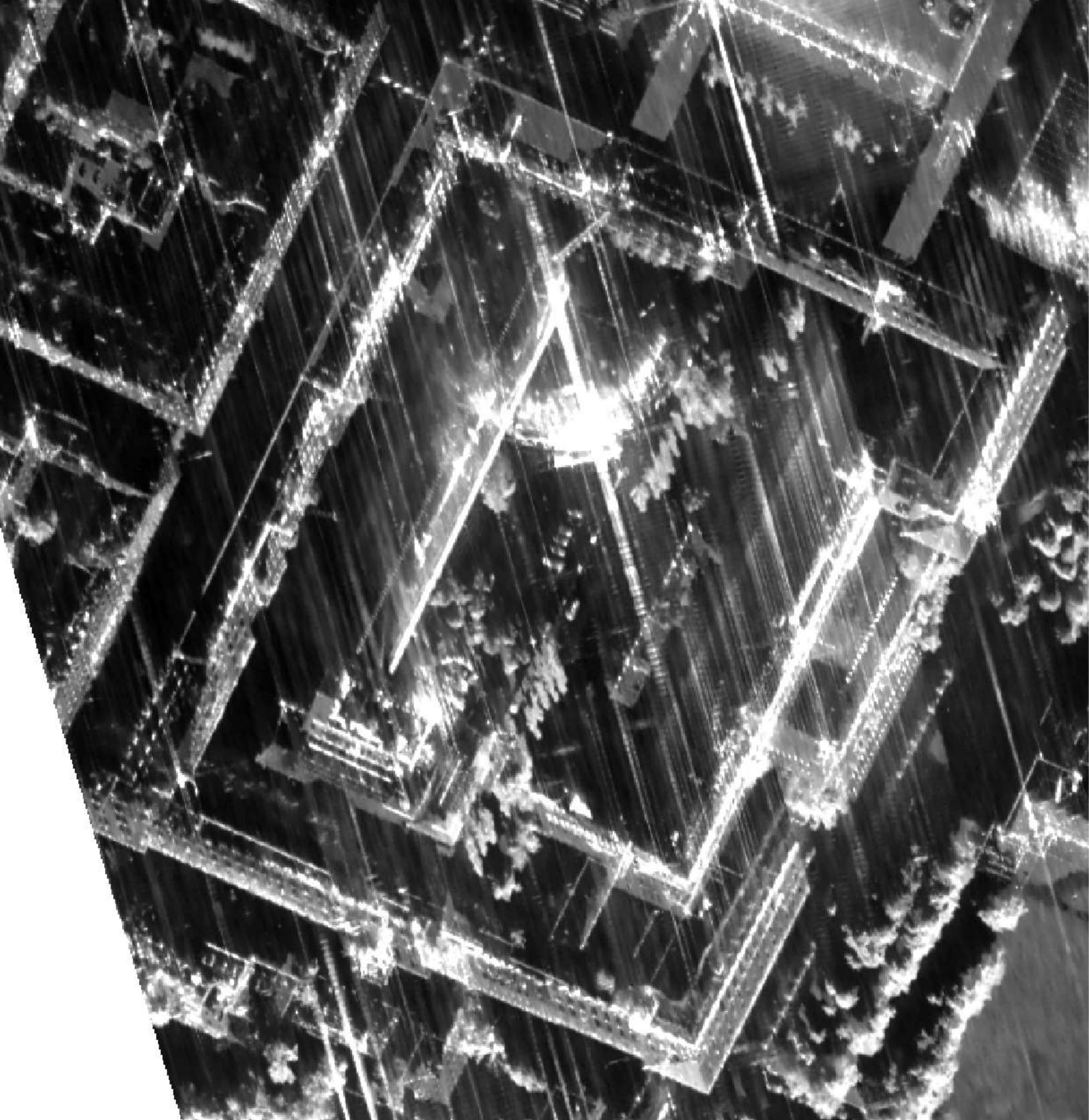}}
 		\subfigure[\textcolor{black}{TSX patch}]{\includegraphics[height=4.3cm]{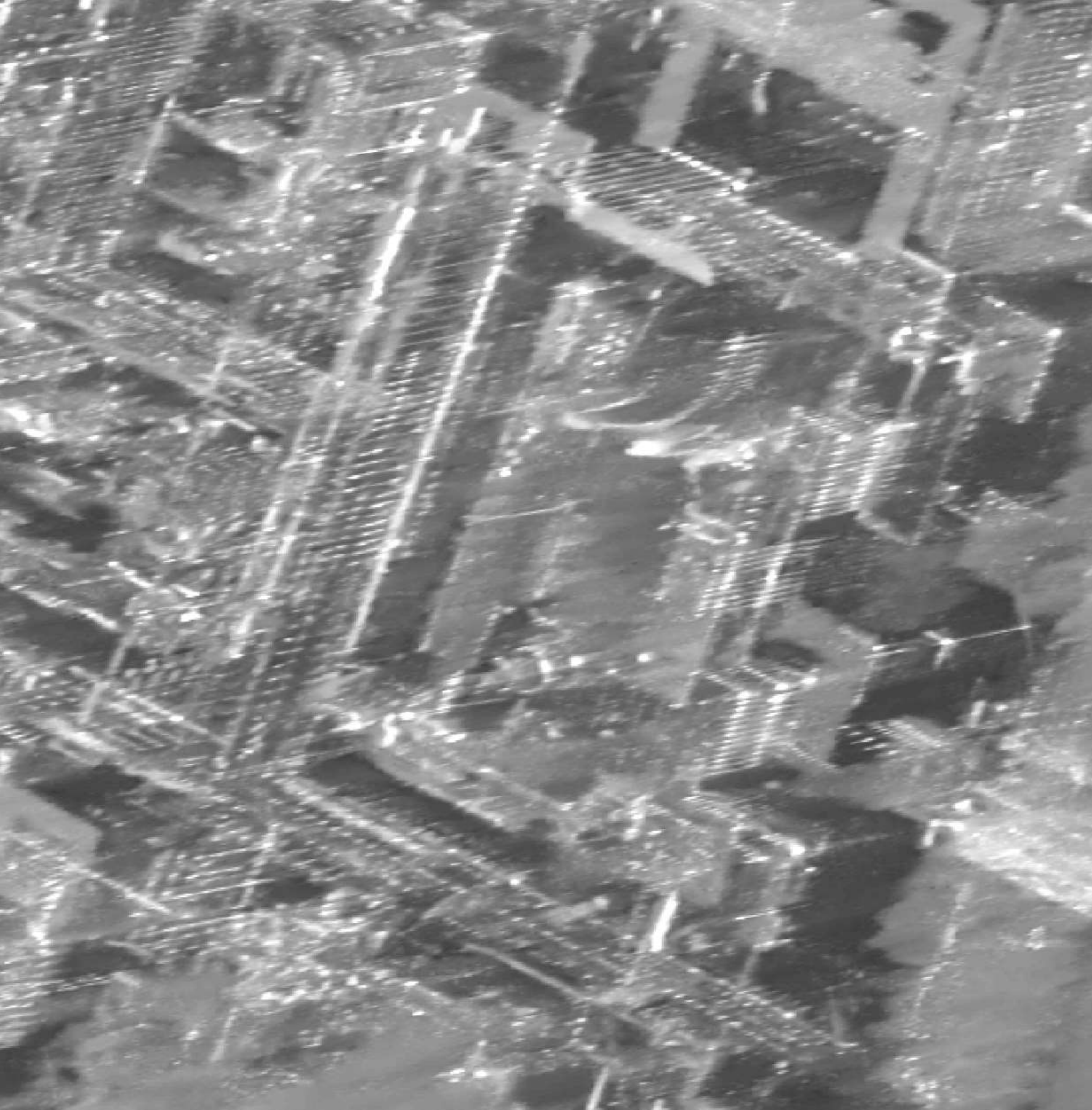}} %sarPatch_air
         	\subfigure[\textcolor{black}{WV2 patch}]{\includegraphics[height=4.3cm]{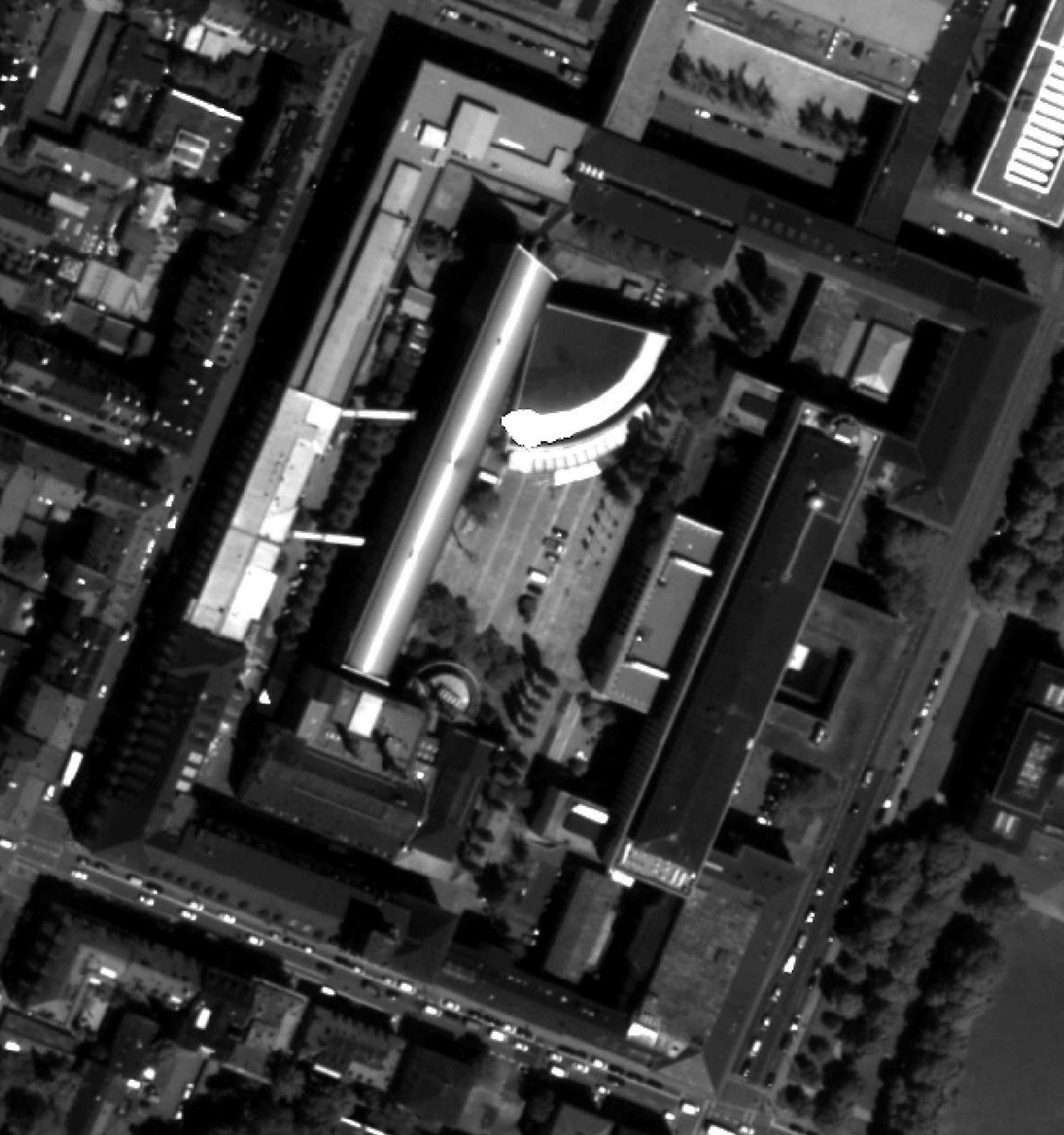}}
 	\caption{Image subsets depicting the TUM main campus for illustration of the effect of pre-processing.}
 	\label{fig:imagePatch}
 \end{figure}

\begin{figure}[htb]  
 	\centering
 	\subfigure[\textcolor{black}{TSX patch}]{\includegraphics[height=4.5cm]{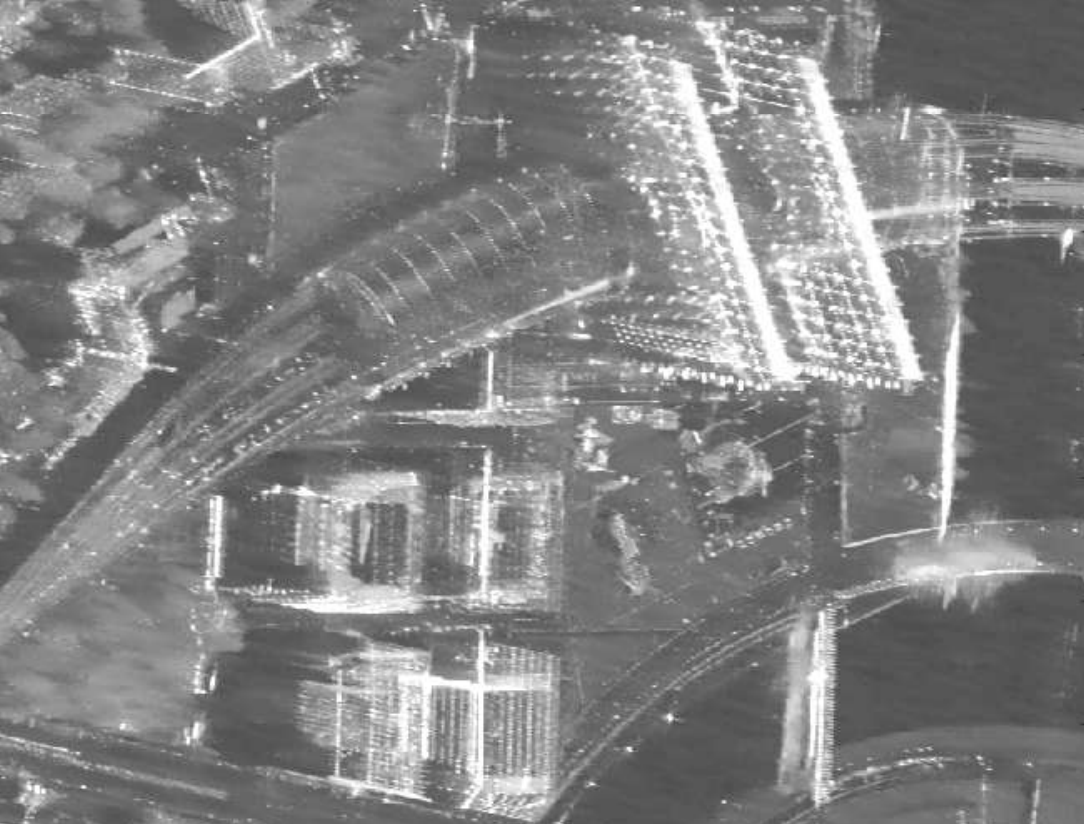}}
 	\subfigure[\textcolor{black}{WV2 patch}]{\includegraphics[height=4.5cm]{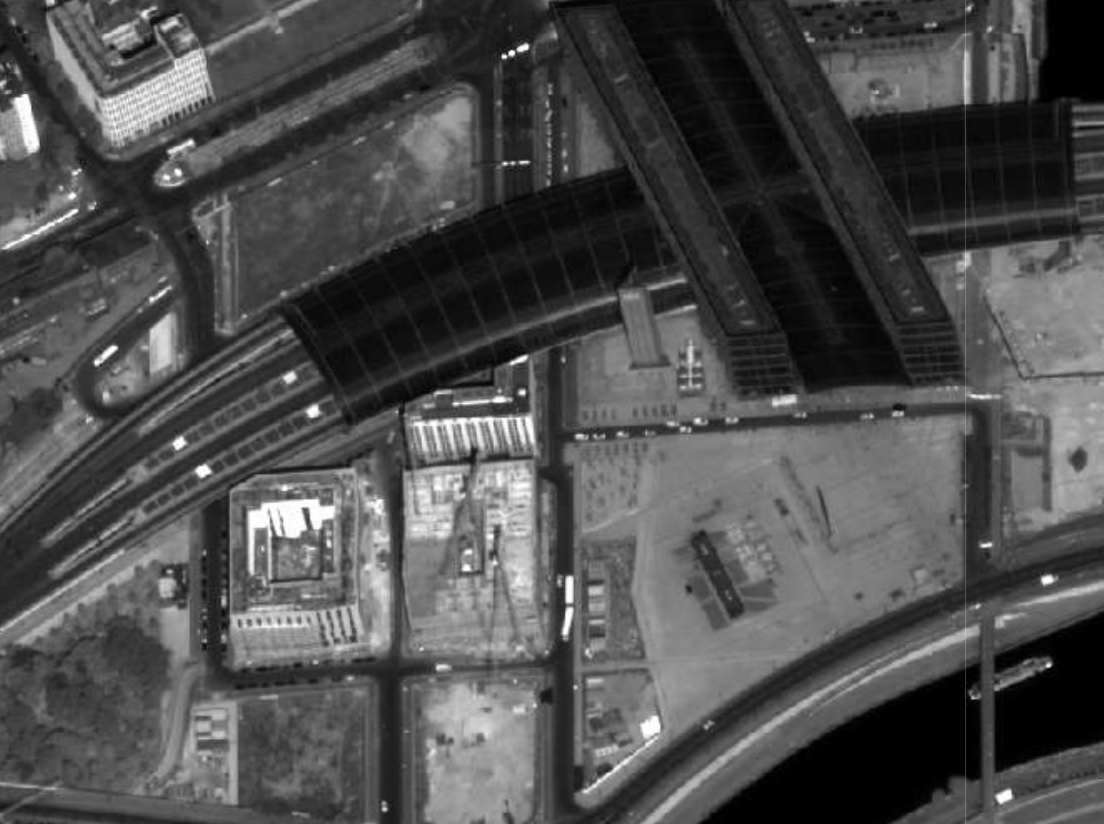}}
 	\caption{Image subsets depicting the main train station of Berlin for illustration of the effect of pre-processing.}
 	\label{fig:imagePatchBerlin}
 \end{figure}

\textcolor{black}{Besides, in order to make the images as similar as possible and ensure the success of tie points selection, all images have been preprocessed to show comparably large, approximately squared pixels (with a pixel spacing of $\sim0.49m $) and a similar, approximately north-aligned orientation. For this purpose, all the SAR images have been firstly filtered nonlocally to mitigate the speckle effect while preserving as many fine details as possible \cite{Deledalle2015}. Afterwards, the resulting despeckled amplitude images were transfered to dB and reduced to approximately square pixels. In addition the SAR images were rotated to an approximate north-aligned orientation. More details about MEMPHIS data preprocessing can be found in \cite{Schmitt2014}. To illustrate the effect of the preprocessing, subsets showing the TUM main campus over Munich and main train station of Berlin are depicted in Fig.~\ref{fig:imagePatch} and  Fig.~\ref{fig:imagePatchBerlin}, respectively.}

For a quantitative evaluation of the stereogrammetric reconstruction results, we used dense LiDAR reference point clouds of centimeter and decimeter accuracy in Munich and Berlin, respectively \cite{Hebel2007, rossi2013urban}.

 \subsection{Experimental Setup}
%\textit{Generic: Here, we will describe which experiments we have carried out.
  
For an investigation of the suitability of the matching strategy proposed in this paper, the five similarity measures described in Section~\ref{sec:SimilarityMeasures} were used. After key point detection, the procedure described in Section~\ref{sec:Matching} was carried out, using SRTM \cite{jarvis2008hole} as weak prior knowledge about the scene topography.

Based on literature knowledge about the vertical accuracy of the SRTM DEM \cite{mukherjee2013evaluation}, the height interval for constructing the IMBLS window was set to $[h_0 -5m,  h_{0} + 20m]$, where $h_{0}$ was taken from the SRTM DEM of the study area. A  $\pm 1$ pixel pre-defined buffer in the row direction was used to form the final IMBLS search window. \textcolor{black}{Based on the empirical finding that larger patches provide better matching results with only little further improvement beyond about $200 \times 200$ pixels, the template size for calculation of the similarity measures was set to $221\times 221$ pixels.} %\todo[inline]{Wow, so big? I thought it was a little smaller, e.g. about 100x100 pixels...} \todo[inline]{\textcolor{black}{Qiu: This is what we use }}

\textcolor{black}{Taking the different original pixel spacings of the three SAR datasets and thus the different information contents within a pre-processed square pixel into consideration, $D_{outlier}< \{3,15,10\}$ pixels was used for the experiments WV2+MEMPHIS (Munich), WV2+TSX (Munich), and WV2+TSX (Berlin), respectively. The rationale behind this choice is simple: when less information is contained in a pixel-defined area, the threshold is chosen accordingly larger.}
  
For quantitative evaluation of the stereogrammetic 3D reconstruction result, point distances to a dense LiDAR reference point cloud were analyzed. To avoid biased results caused by mismatches between the reconstructed 3D points and the LiDAR data, the distance is not calculated based on individual point neighbors, but on a plane fitted in a least-squares sense through the 10 nearest neighbors \cite{Schmitt2014}.

\subsection{Tie Point Matching Results} \label{sec:MatchingResults}
\begin{figure}[H]%[!tbh]
	\centering
	
	\subfigure[SAR image.]{
		%	\rule{4cm}{3cm}
		\includegraphics[height=6.5cm]{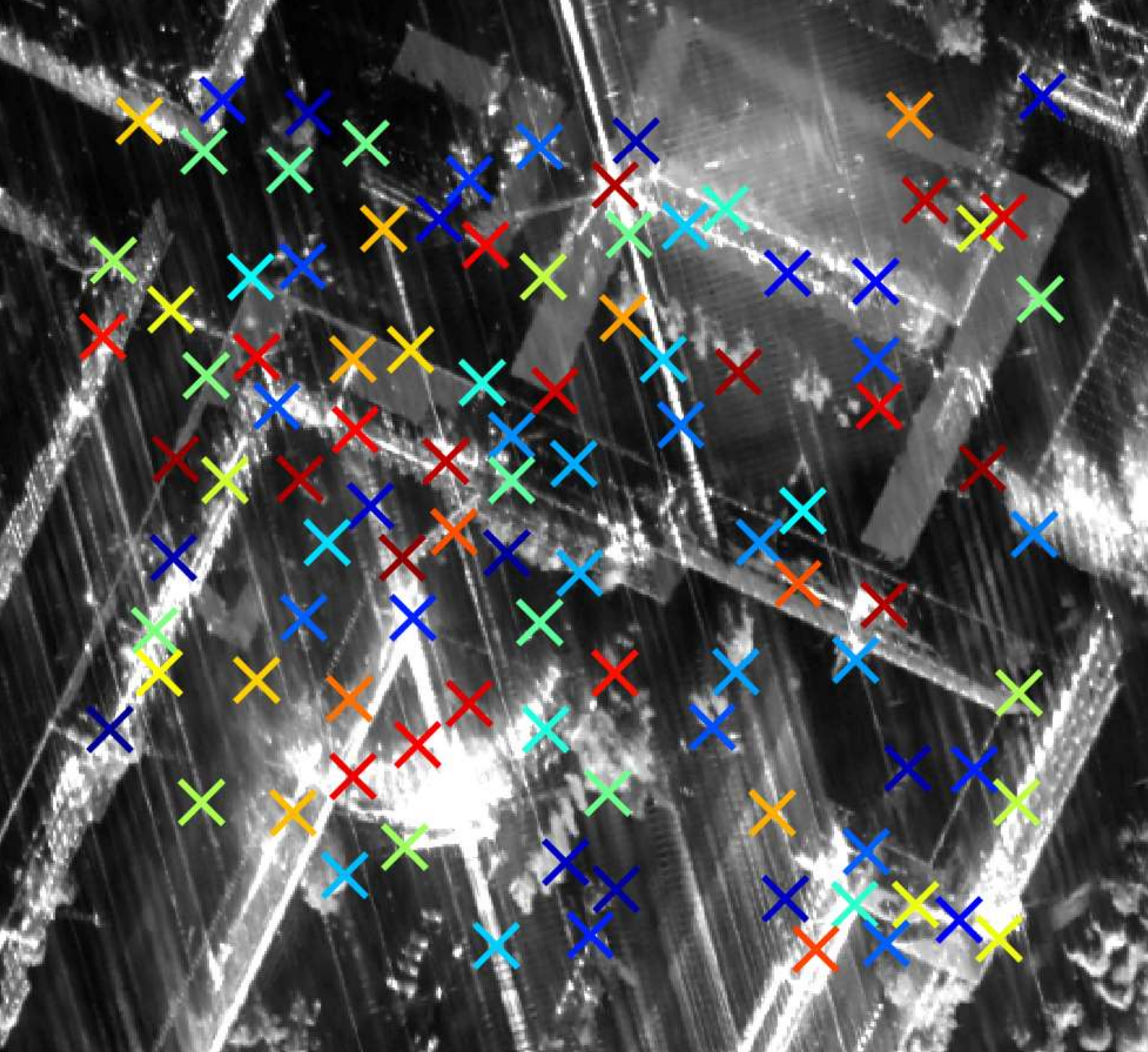}
		\label{}
	}%
	\subfigure[Optical image.]{
		%	\rule{4cm}{3cm}
		\includegraphics[height=6.5cm]{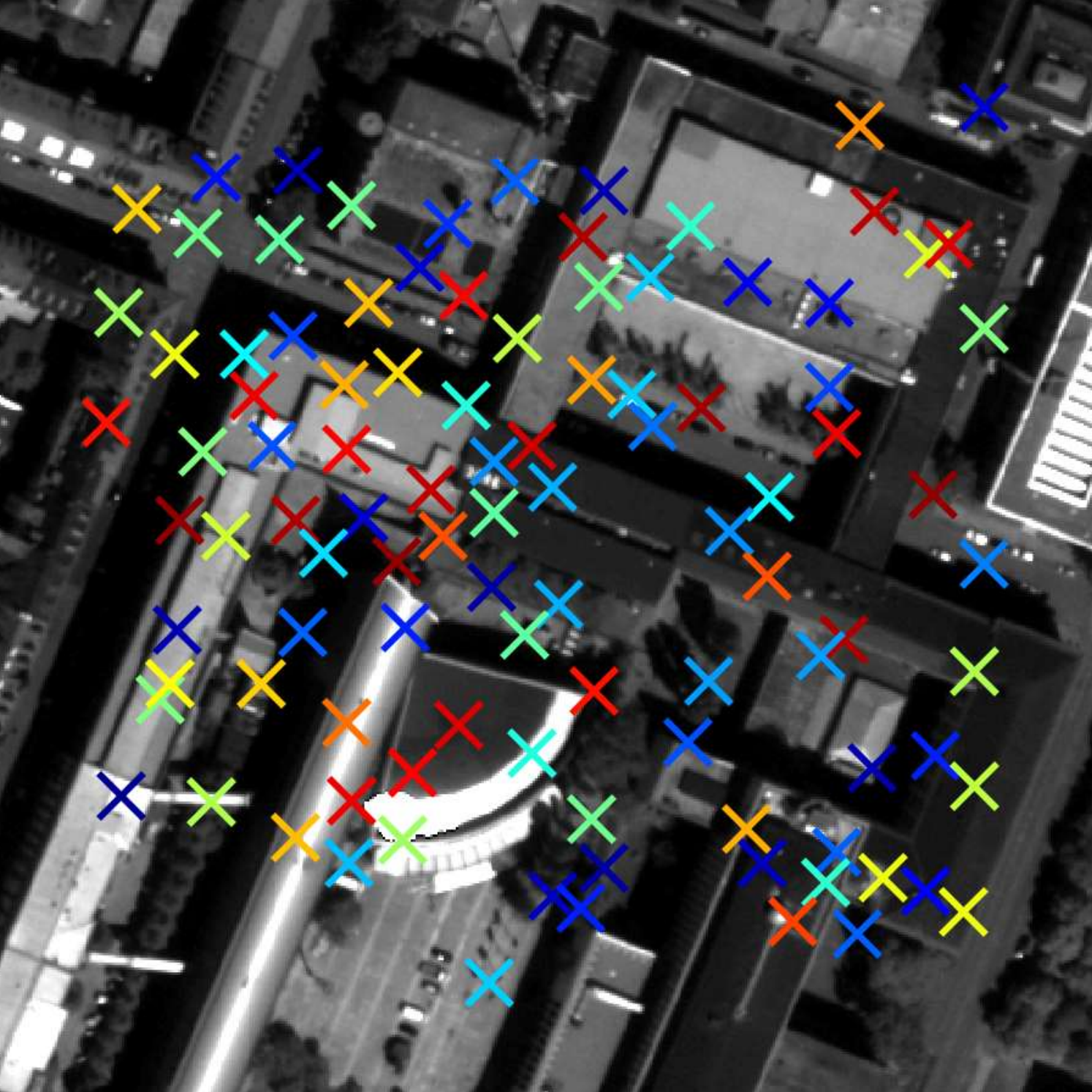}
		\label{}
	}
	\caption{Tie points selection result using HOPC and data WV2+MEMPHIS over Munich. 102 out of 739 tie points are shown; The color is coupled only to show the correspondences. }
	\label{fig:tieOriginal_air}
\end{figure}

\begin{figure}[H]
	\centering
	
	\subfigure[SAR image.]{
		%	\rule{4cm}{3cm}
		\includegraphics[height=6.0cm]{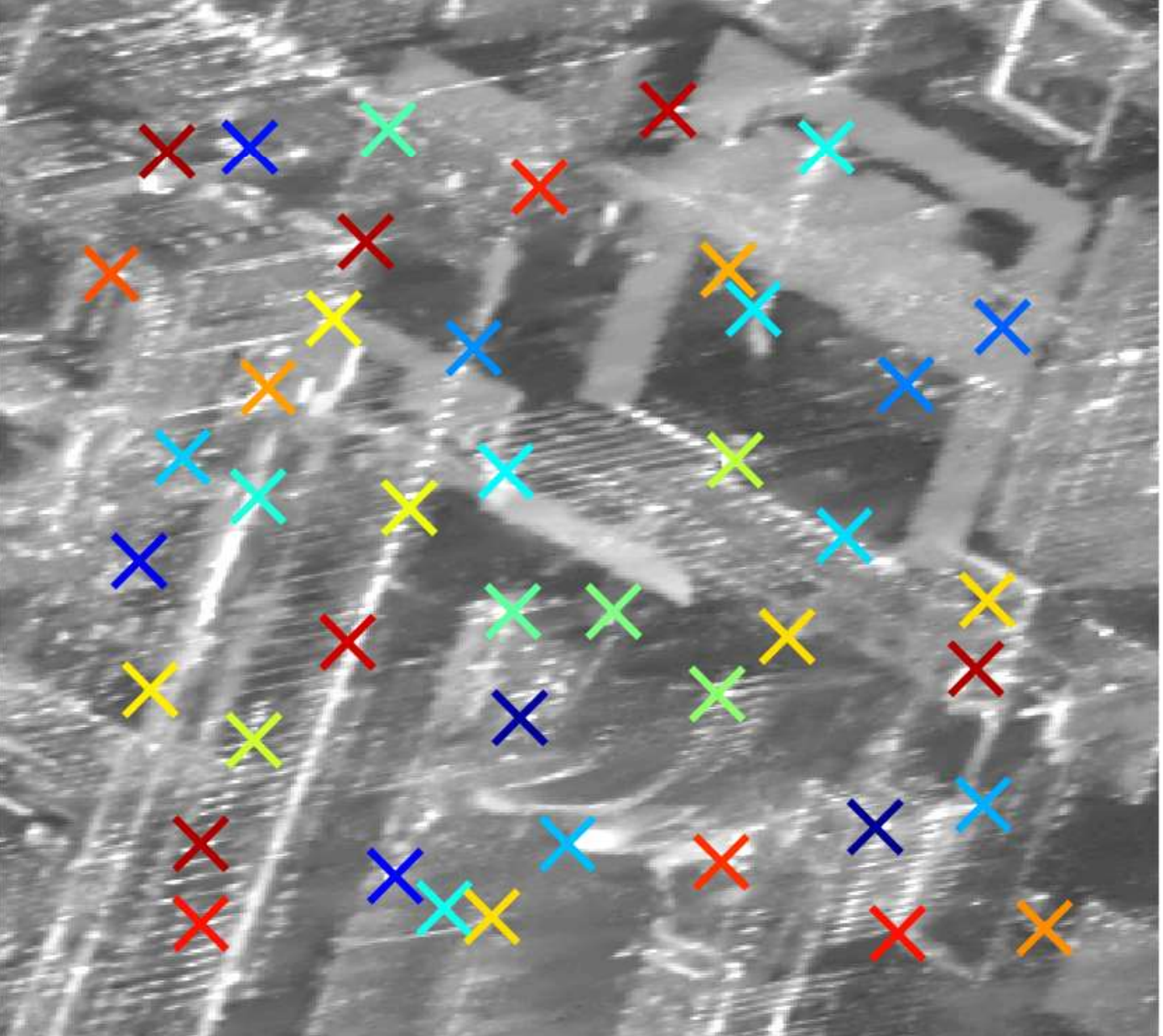}
		\label{}
	}%
	\subfigure[Optical image.]{
		%	\rule{4cm}{3cm}
		\includegraphics[height=6.0cm]{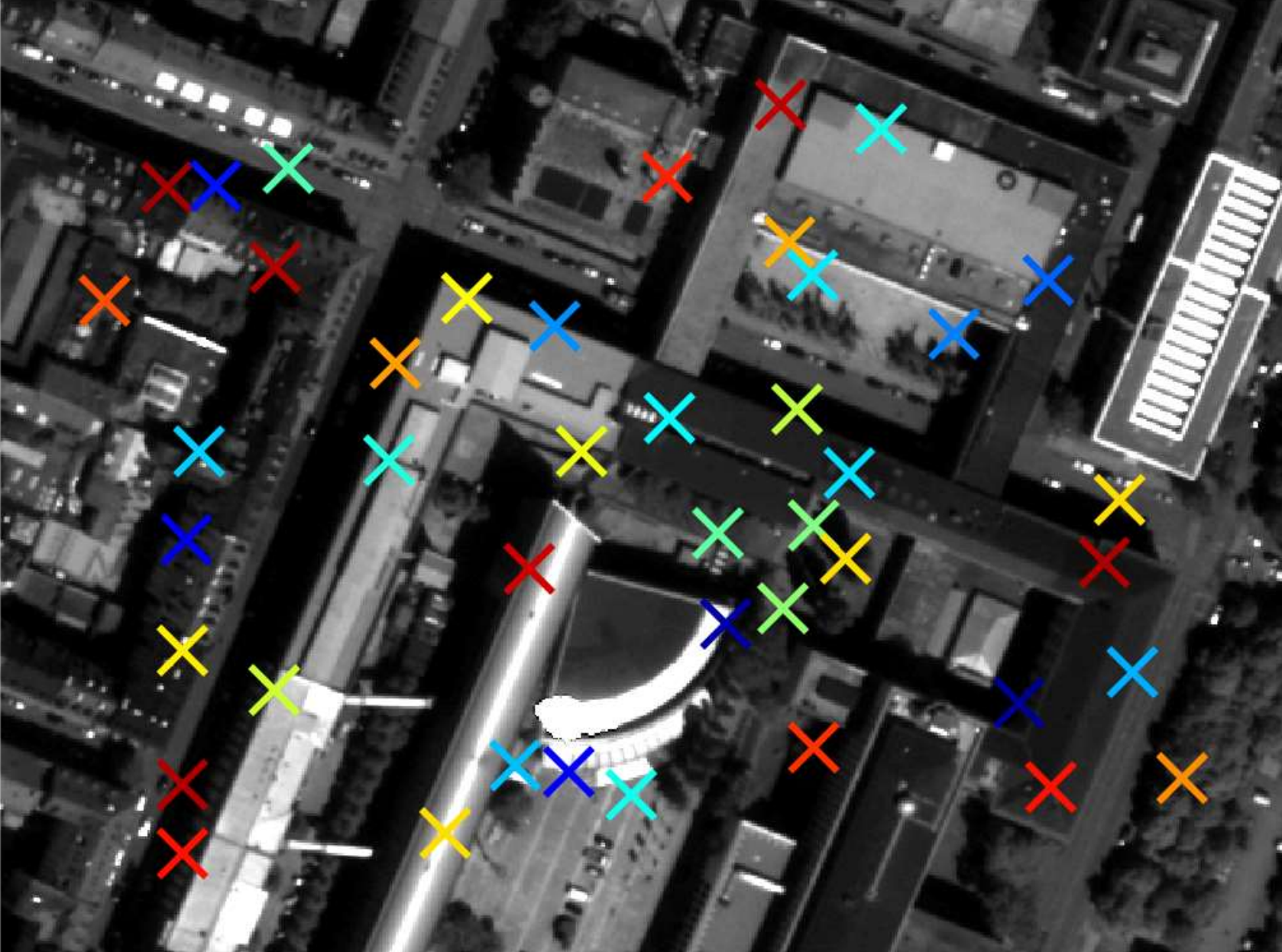}
		\label{}
	}
	\caption{Tie points selection result using HOPC and data WV2+TSX over Munich. 43 out of 699 tie points are shown; The color is coupled only to show the correspondences. }
	\label{fig:tieOriginal}
\end{figure}

\begin{figure}[H]
	\centering
	
	\subfigure[SAR image.]{
		%	\rule{4cm}{3cm}
		\includegraphics[height=6.5cm]{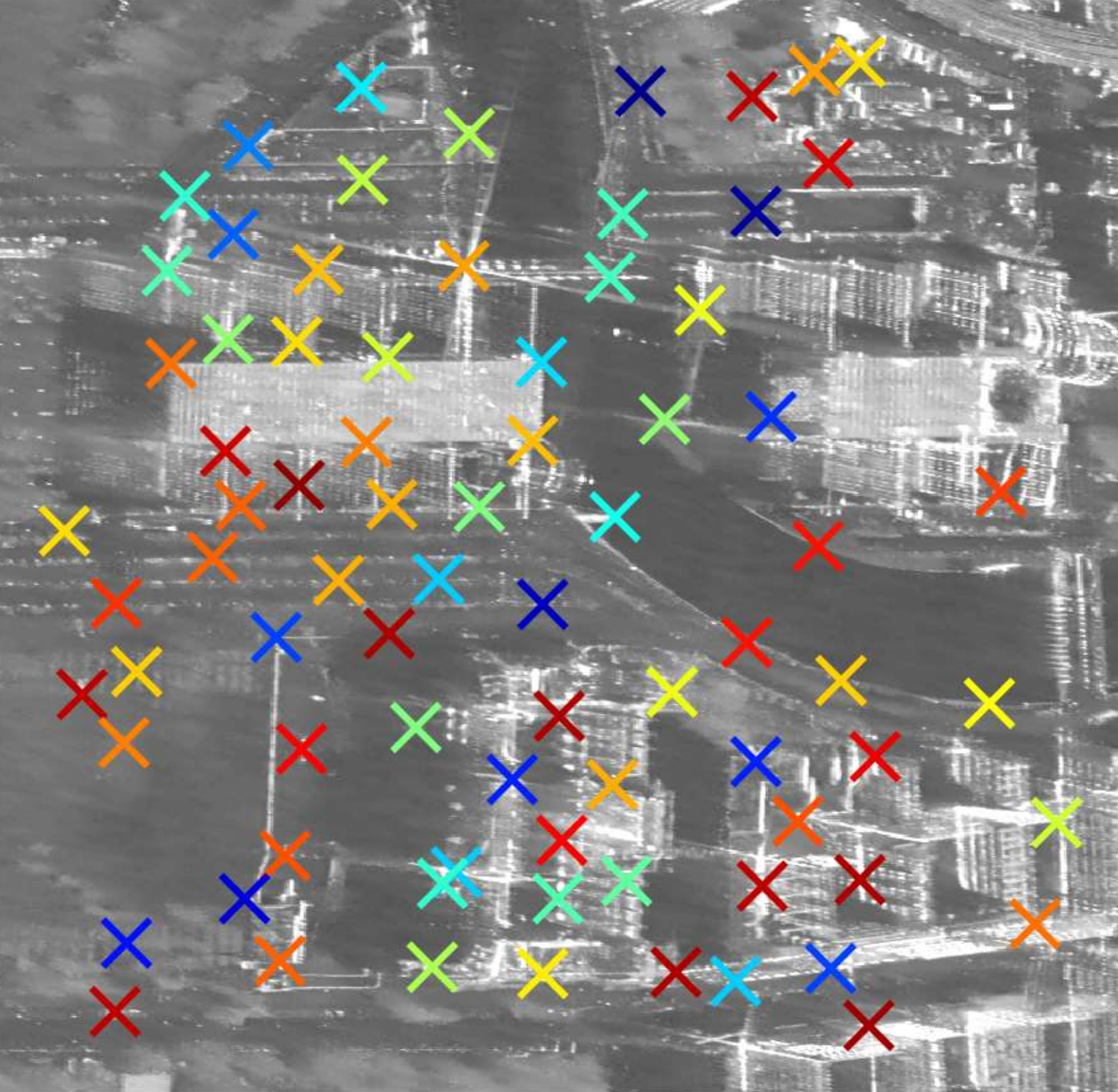}
		\label{}
	}%
	\subfigure[Optical image.]{
		%	\rule{4cm}{3cm}
		\includegraphics[height=6.5cm]{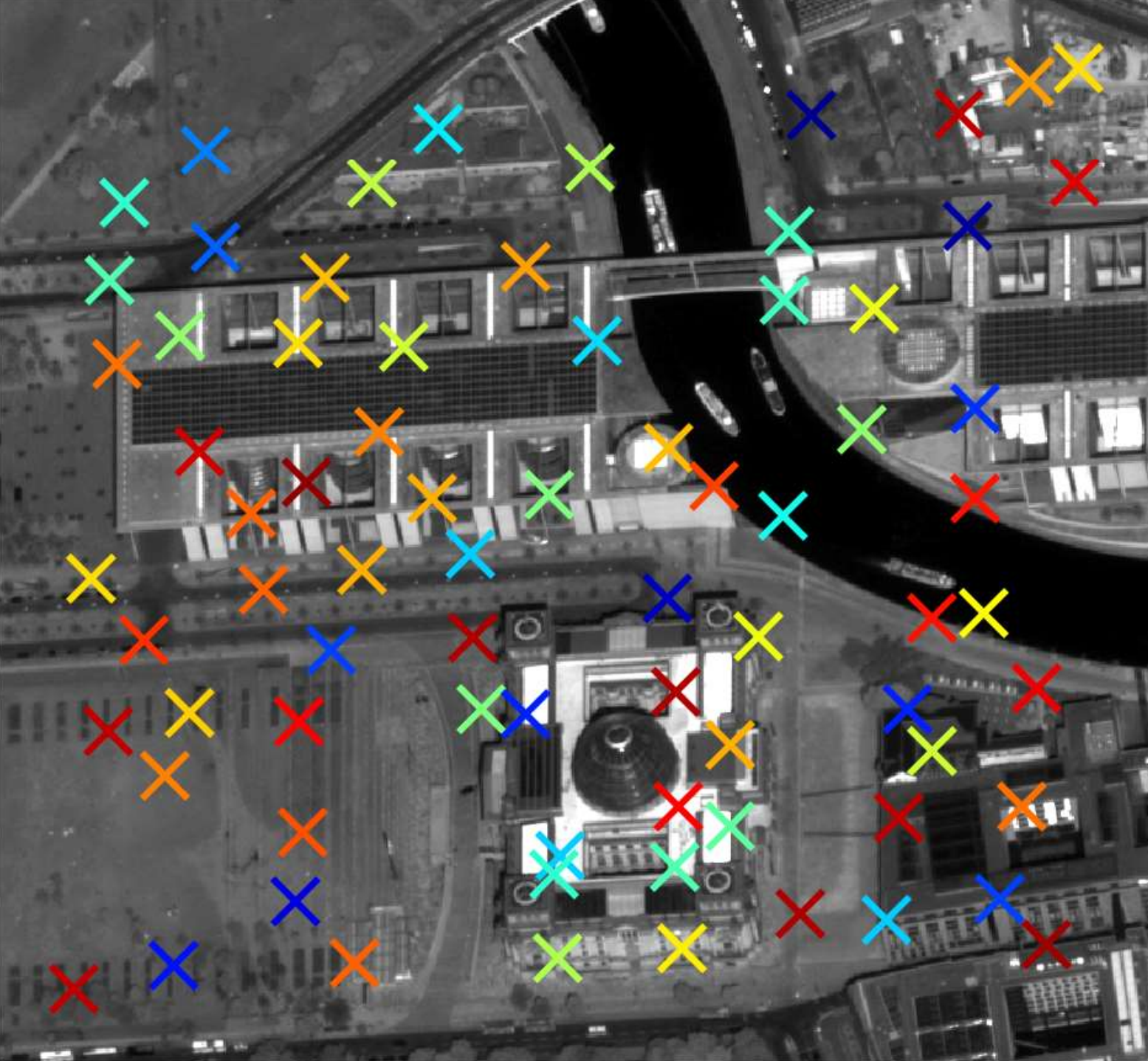}
		\label{}
	}
	\caption{Tie points selection result using HOPC and data WV2+TSX over Berlin. 78 out of 446 tie points are shown; The color is coupled only to show the correspondences. }
	\label{fig:tieAirOriginal_berlin}
\end{figure}

%\subsection{Tie Point Matching Results using the IMBLS window} \label{sec:IMBLSMatchingResults}	
\textcolor{black}{The matching results relying on HOPC as similarity measure and the IMBLS search strategy without subsequent outlier removal are exemplarily displayed in Fig.~\ref{fig:tieOriginal_air} for WV2+MEMPHIS over Munich, Fig.~\ref{fig:tieOriginal} for WV2+TSX over Munich and Fig.~\ref{fig:tieAirOriginal_berlin} for WV2+TSX over Berlin. Considering conciseness and clarity, only a subset of the corresponding experimental scene is shown.}

\begin{figure}[H]
	\centering
    
	\subfigure[SAR image.]{
		%	\rule{4cm}{3cm}
		\includegraphics[height=6.5cm]{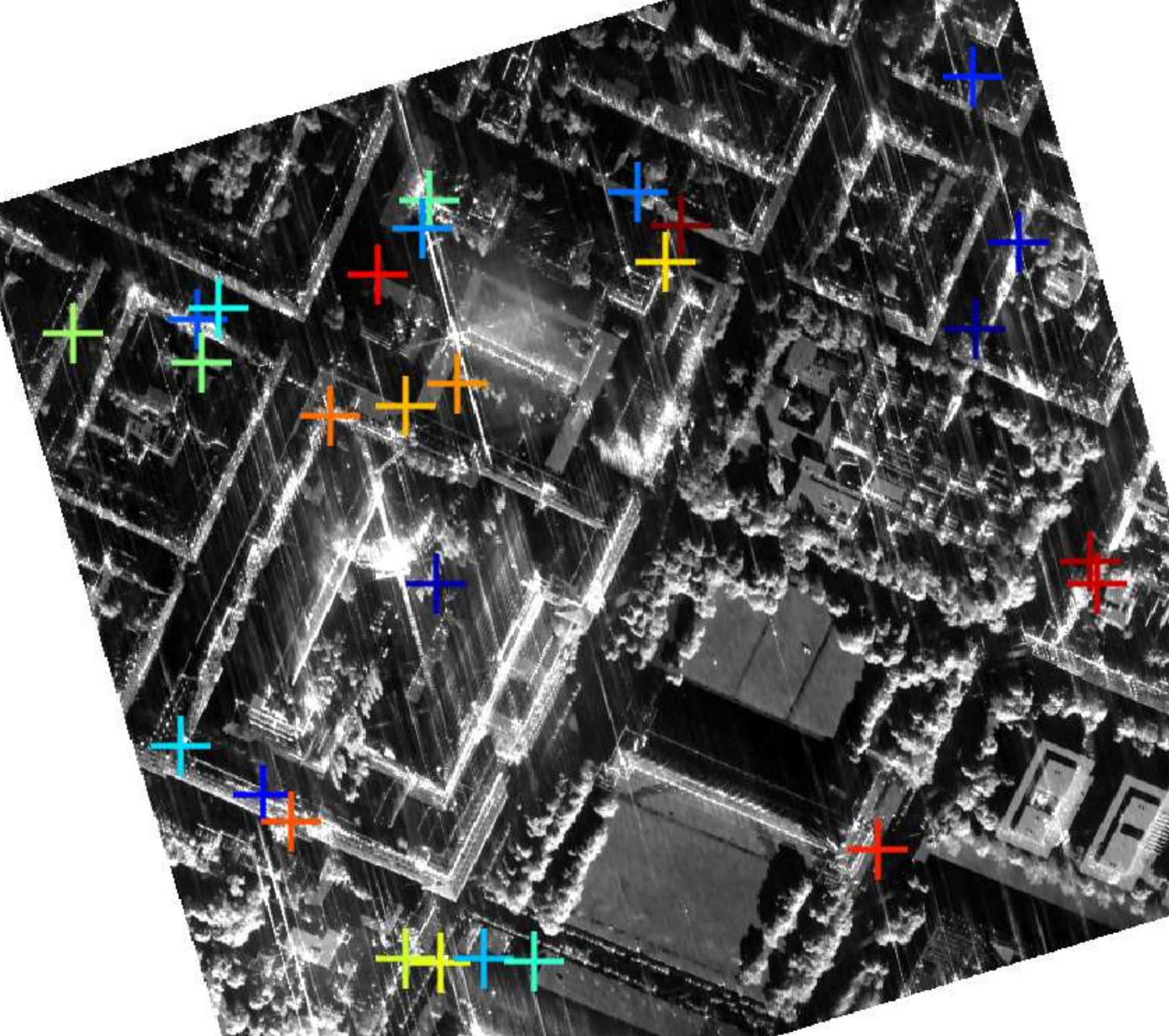}
		\label{}
	}%
	\subfigure[Optical image.]{
		%	\rule{4cm}{3cm}
		\includegraphics[height=6.5cm]{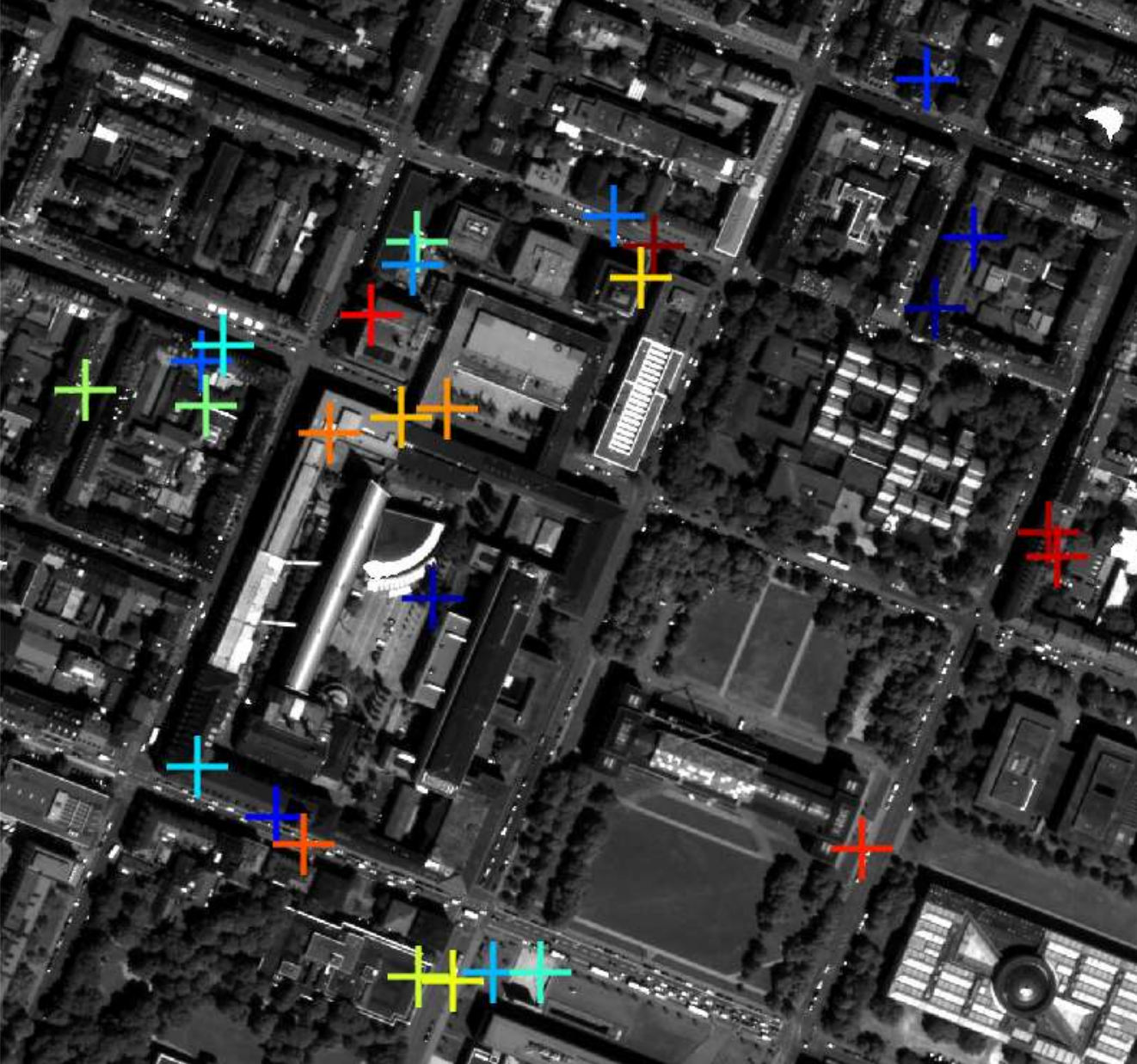}
		\label{}
	}
	\caption{Tie points selection result using HOPC and data WV2+MEMPHIS over Munich. Number of points: 27; The color is coupled only to show the correspondences.}
	\label{fig:tie_air_outlierRemove}
\end{figure}

\begin{figure}[H]
	\centering
    
	\subfigure[SAR image.]{
		%	\rule{4cm}{3cm}
		\includegraphics[height=7.5cm]{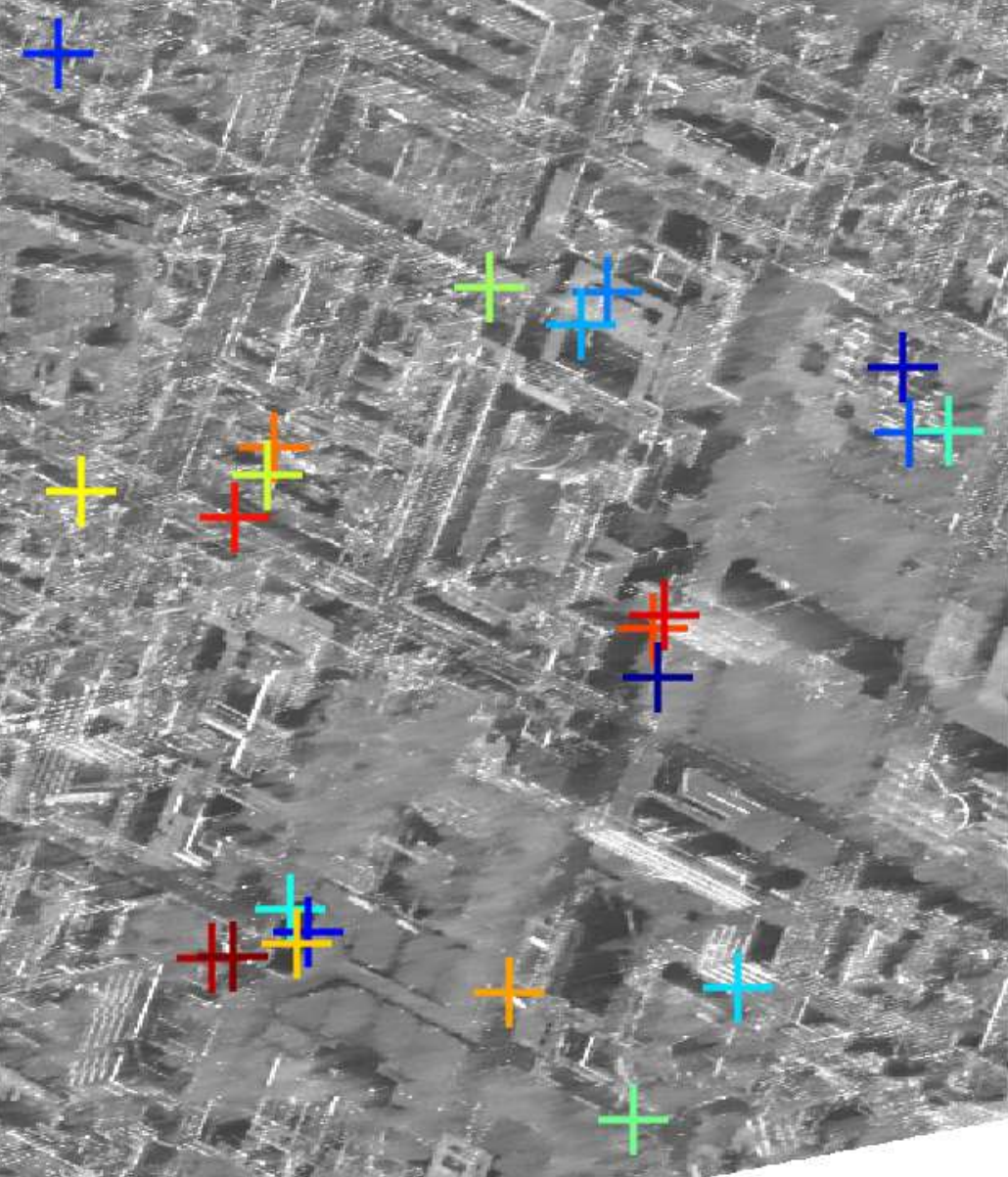}
		\label{}
	}%
	\subfigure[Optical image.]{
		%	\rule{4cm}{3cm}
		\includegraphics[height=7.5cm]{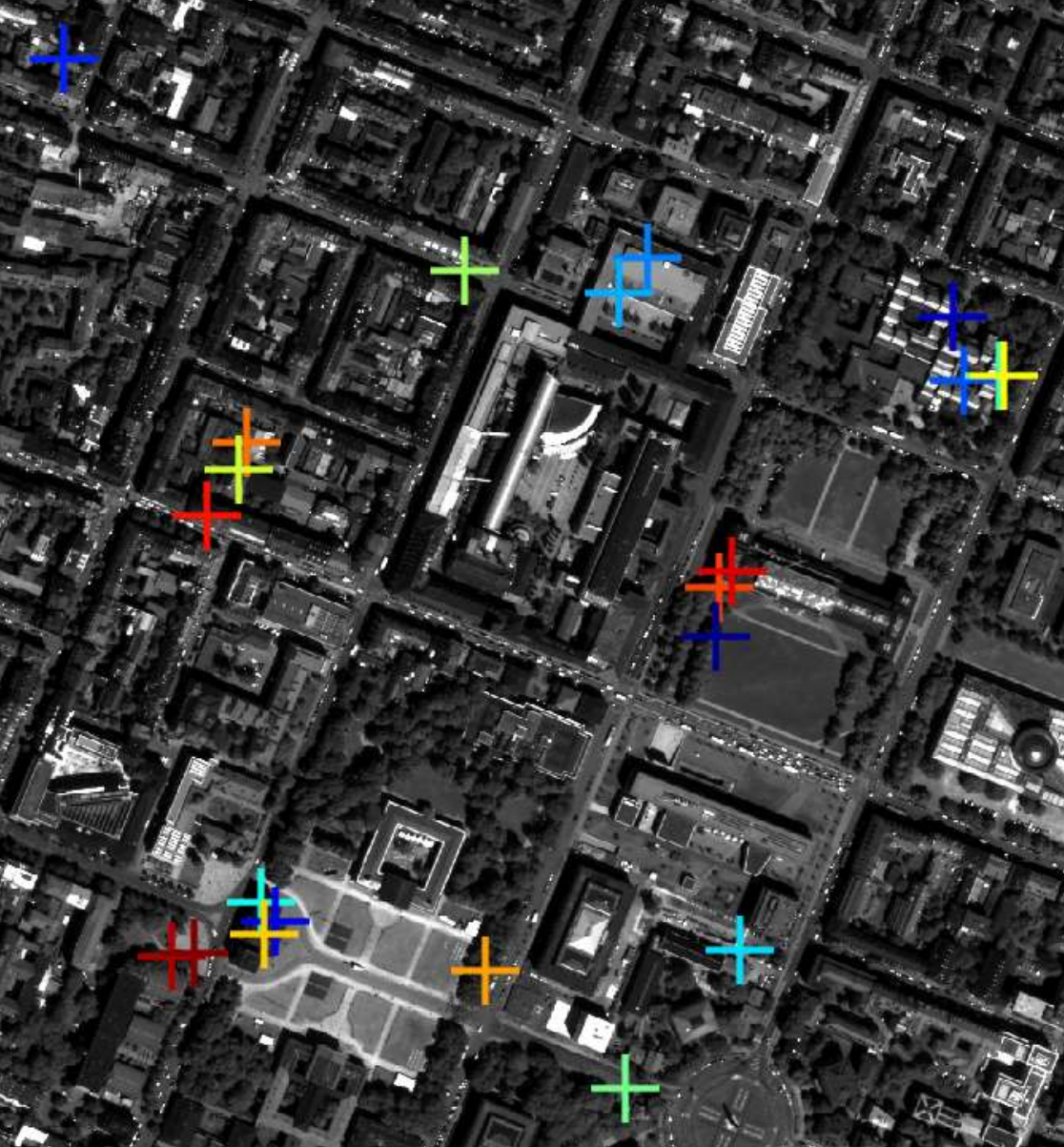}
		\label{}
	}
	\caption{Tie points selection result using HOPC and data WV2+TSX over Munich. Number of points: 22; The color is coupled only to show the correspondences. }
	\label{fig:tie_outlierRemove}
\end{figure}

\begin{figure}[H]
	\centering
    
	\subfigure[SAR image.]{
		%	\rule{4cm}{3cm}
		\includegraphics[height=5.4cm]{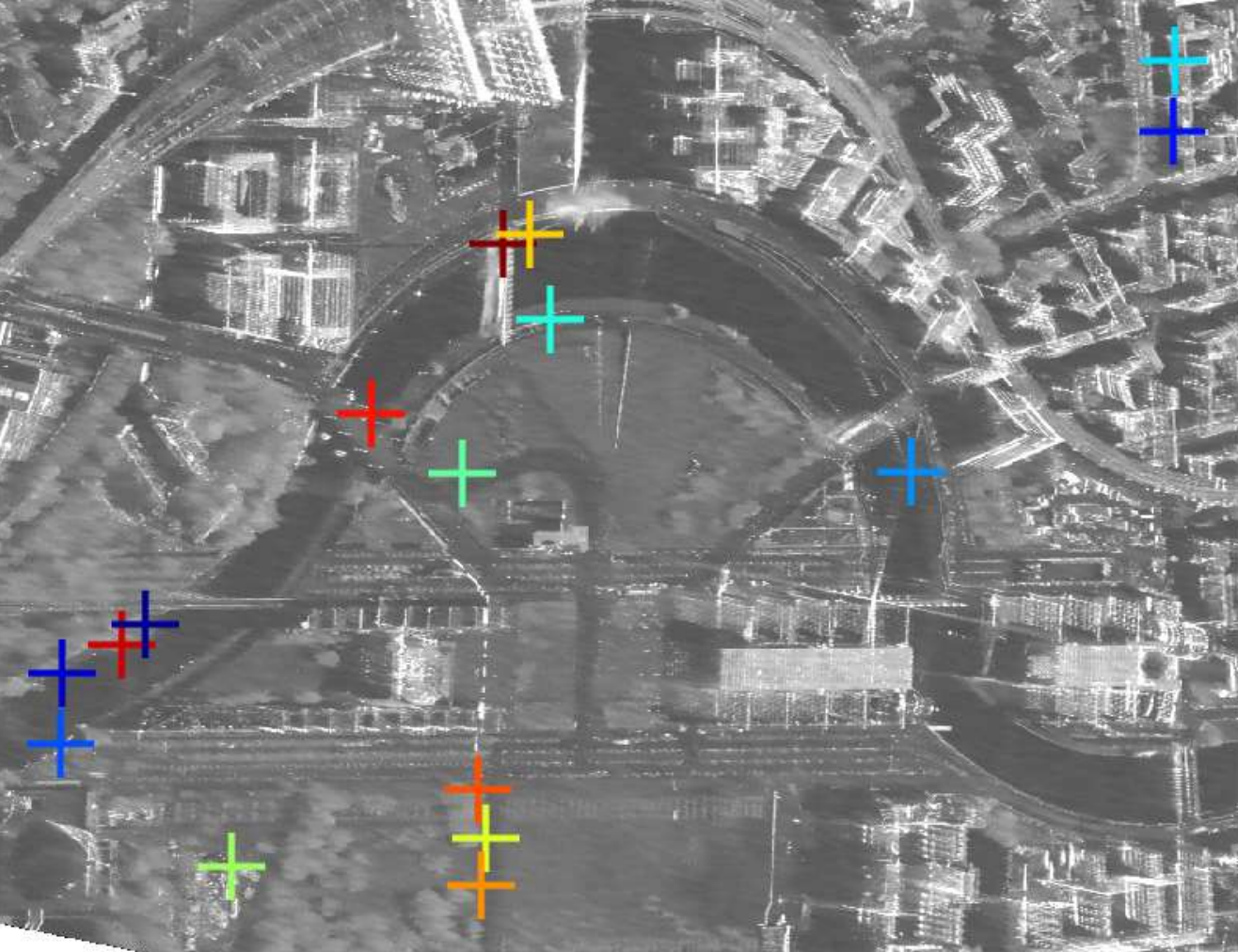}
		\label{}
	}%
	\subfigure[Optical image.]{
		%	\rule{4cm}{3cm}
		\includegraphics[height=5.4cm]{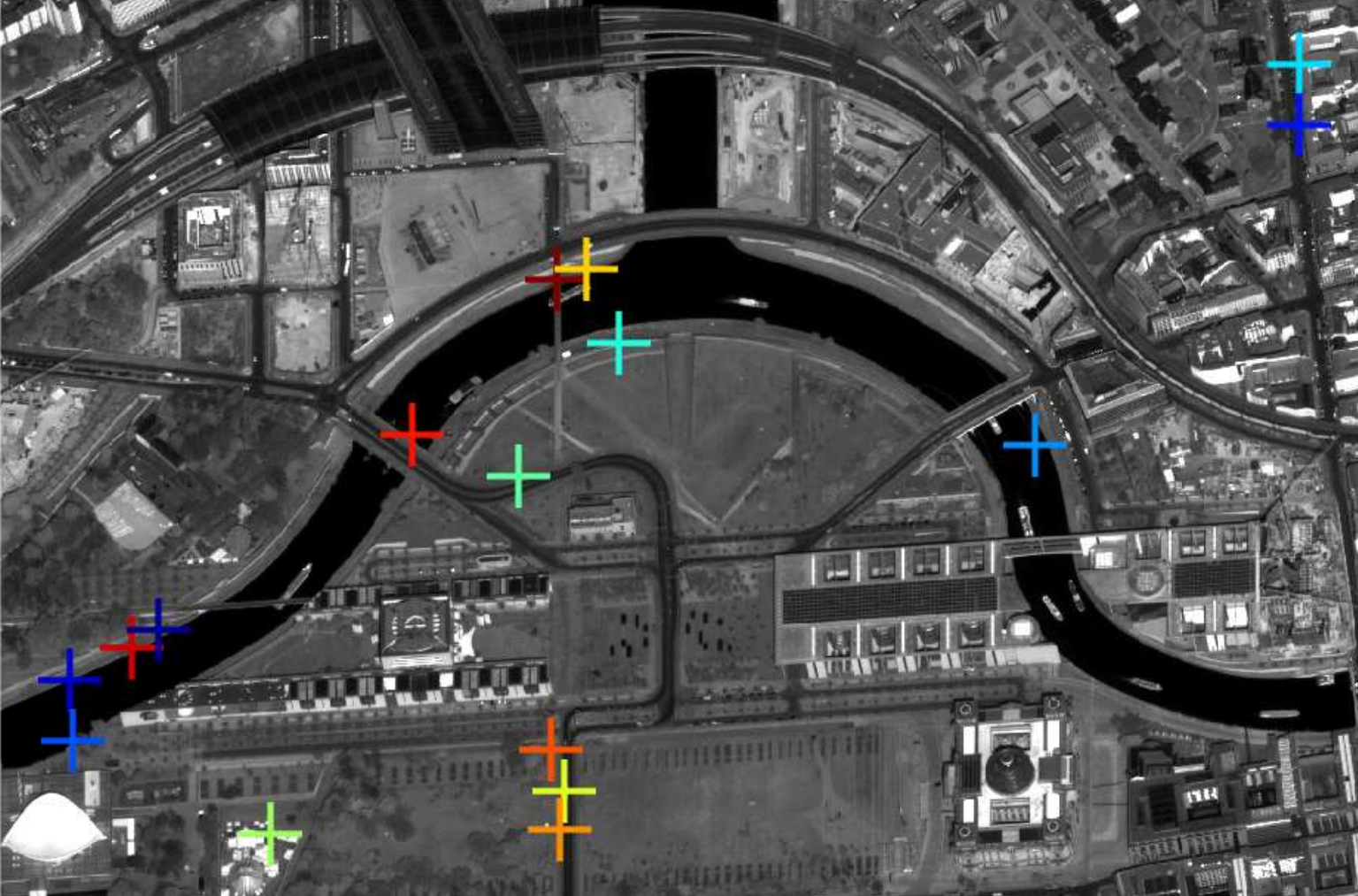}
		\label{}
	}
	\caption{Tie points selection result using HOPC and data WV2+TSX over Berlin. Number of points: 16; The color is coupled only to show the correspondences. }
	\label{fig:tie_berlin_outlierRemove}
\end{figure}

   %  \subsection{Tie Point Matching Results after outliers removal} \label{sec:MatchingResultsAfterOutlierexlusion}	   
In analogy, the results achieved using the outlier removal approach exploiting all similarity measures in a joint manner can be seen in Figs.~\ref{fig:tie_air_outlierRemove},~\ref{fig:tie_outlierRemove} and~\ref{fig:tie_berlin_outlierRemove}, respectively.

\subsection{3D-Reconstruction Results} \label{sec:StereoResults}

To evaluate the quality of the matching process, the results of the stereogrammetric 3D reconstruction can be found in Tab.~\ref{tab:statisReconstruction}, 
which shows the number of the resulting points, \textcolor{black}{the mean, mean square error (MSE) and median of the absolute distances (in m) as well as the share of points with a distance to the nearest least-squares plane below one meter (percentage). For sake of comparison, Mean and Standard Deviation (SD) of results from different similarity measures are also shown}, while the last row indicates the heights of the key points, resulting from single image positioning with assistance of the SRTM DEM.

\begin{table}[H]
	\newcommand{\tabincell}[2]{\begin{tabular}{@{}#1@{}}#2\end{tabular}}
	\centering
\caption{Quantitative evaluation of the stereogrammetric 3D reconstruction results}
 \begin{tabular} {p{2.2cm}<{\centering}p{1.8cm}<{\centering}p{1.4cm}<{\centering}p{2cm}<{\centering}p{2cm}<{\centering}p{2.5cm}<{\centering}}
	\toprule
	\multirow{2}[2]{*}{data} & \multirow{2}[2]{*}{\tabincell{c}{similarity  \\ measure}}
	& \multicolumn{3}{c}{distance to nearest plane (in m)} & \multirow{2}[2]{*}{ \tabincell{c}{percentage}      } \\
	&       &mean     &MSE  & median&  \\
	\midrule
	\multirow{6}[6]{*}{\tabincell{c}{Munich,\\ WV2\\ +MEMPHIS\\(27 points)}}
    & NCC    &    1.88  & 2.79  & 1.49  & 41\% \\
   & MI &  {1.78} & 2.69  & 1.25  & 41\% \\
    & HOG    & 1.90  & 2.93  & 1.05  & 48\% \\
    & SIFT  &  1.82  & 2.93  & 1.10  & 44\% \\
    & HOPC  &   1.79  &  {2.65} & 1.12  & 37\% \\
 
   \cmidrule{2-6}   
   &  {mean}   &       \textbf{1.84} & \textbf{2.80} & \textbf{1.20} & \textbf{42\%} \\
    &  {SD}   &         \textbf{0.05} & \textbf{0.13} & \textbf{0.18} & \textbf{0.04} \\

   \cmidrule{2-6}   
   & \textit{SRTM}   & \textit{2.22}   & \textit{2.90}  & \textit{1.65}  & \textit{30\%} \\
   
	\midrule
	\multirow{6}[6]{*}{\tabincell{c}{Munich,\\ WV2\\ +TSX\\(22 points)}  
	}
    & NCC  & 1.86  & 2.76  & 1.11 & 41\% \\
    & MI &   1.72  & 2.71  & 1.26  & 45\% \\
    & HOG    &  1.92  & 2.74  & 1.54  & 36\% \\
    & SIFT  &  1.49  & 2.49  & 0.86   & 55\% \\
    & HOPC  &    {1.49} &  {2.20} & 0.96 & 55\% \\
    
   \cmidrule{2-6}   
   & \textit{mean}   & \textbf{1.70} & \textbf{2.58} & \textbf{1.15} & \textbf{46\%} \\
       &  {SD}      &  \textbf{0.20} & \textbf{0.24} & \textbf{0.27} & \textbf{0.08} \\

   \cmidrule{2-6}   
   & \textit{SRTM}   &      \textit{3.75} & \textit{5.90} & \textit{3.39} & \textit{23\%} \\

    	\midrule
	\multirow{6}[6]{*}{\tabincell{c}{Berlin,\\ WV2\\ +TSX\\(16 points)}  
	}
   & NCC  &   2.00  & 2.77  & 1.14  & 50\% \\
   & MI &  2.01  & 2.69  & 1.33  & 44\% \\
   & HOG    &   1.83  & 2.45  & 1.39  & 38\% \\
   & SIFT  &     1.92  & 2.79  & 1.26  & 44\% \\
   & HOPC  &    {1.77} &  {2.28} & 1.28  & 31\% \\
   
      \cmidrule{2-6}   
   & \textit{mean}   & 
    \textbf{1.91} & \textbf{2.60} & \textbf{1.28} & \textbf{41\%} \\
    &  {SD}   &     \textbf{0.10} & \textbf{0.22} & \textbf{0.09} & \textbf{0.07} \\
   
   \cmidrule{2-6}   
   & \textit{SRTM}   &  
    \textit{3.18} & \textit{4.46} & \textit{2.46} & \textit{31\%} \\
	\bottomrule
\end{tabular}%
\label{tab:statisReconstruction}%
\end{table}%

\section{Discussion}\label{sec:Discussion}

As the results of the experiments illustrate, \textcolor{black}{in all cases a reasonable number of robustly matched points is found: 27 in the WV2+MEMPHIS case over Munich, 28 in the WV2+TSX case over Munich, and 22 over the WV2+TSX case over Berlin}. Looking at the 3D-reconstruction results, accuracies in the 1-2m domain are achieved, which provides a significant improvement over the prior knowledge provided by SRTM. These results can be interpreted as follows: 

\subsection{Difference Between Airborne and Spaceborne SAR Data}
The difference in finally matched tie points for the MEMPHIS and the TSX cases can be explained by the different appearance of the datasets. While the MEMPHIS image was acquired in Ka-band, and thus contains less speckle noise and more rough-appearing surfaces, as well as an ultra-high resolution in the centimeter-range, the TSX staring spotlight images were acquired in X-band and provides a slightly lower resolution in the decimeter-range.
%higher local similarity due to the high resolution, which shows individual details more clear and similar to WV than TSX. 
Thus, \textcolor{black}{corresponding image patches} from the MEMPHIS image and the WV2 image look more similar \textcolor{black}{locally} than those from the TSX images and the WV2 image, \textcolor{black}{which is also confirmed by the similar performance of all similarity measures, with a smallest standard deviation for mean as well as MSE (cf. Tab.~\ref{tab:statisReconstruction}) in the MEMPHIS case.}
%, as NCC works best for only slightly distorted images. 

\begin{figure*}[!tbh]
	\centering
    
	\subfigure[WV2+MEMPHIS]{
		%	\rule{4cm}{3cm}
		\includegraphics[width=0.31\textwidth]{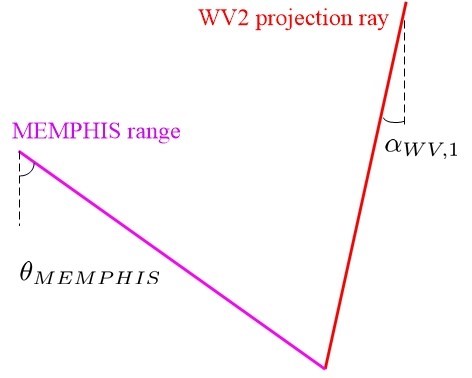}
		\label{}
	}%
	\subfigure[WV2+TSX]{
		%	\rule{4cm}{3cm}
		\includegraphics[width=0.21\textwidth]{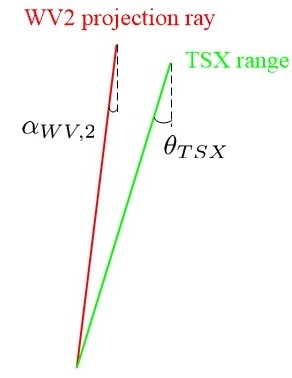}
		\label{}
	}
    \subfigure[WV2+TSX(Berlin)]{
		%	\rule{4cm}{3cm}
		\includegraphics[width=0.23\textwidth]{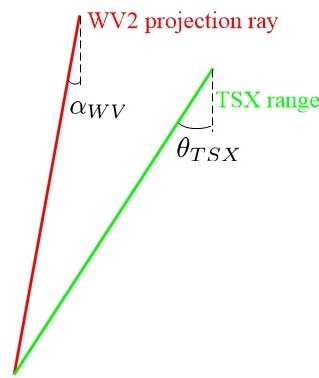}
		\label{}
	}%
	\caption{The stereo geometry of the datasets used in the experiment. The plane is defined by the Range-Doppler circle.}
	\label{fig:interSectionPlane}
\end{figure*}

\textcolor{black}{In contrast to the matching success rate stands the performance achieved during 3D-reconstruction. Here, the WV2+TSX case over Munich provides the best result by a small margin with accuracies of about 2.58m (MSE) and about 46\% of all reconstructed points residing within 1m from the nearest least-squares plane fitted into the LiDAR reference data. The WV2+MEMPHIS case over Munich and the WV2+TSX case over Berlin provide similar MSEs of about 2.80m and 2.60m, but only about 42\% and 41\% of all reconstructed points residing within 1m from the nearest reference plane, respectively. This can be explained by consideration of the intersection geometries derived in Section~\ref{sec:SAROPTprinciple}: As shown in Fig.~\ref{fig:interSectionPlane}, WV2+MEMPHIS follows an opposite-side stereo case, while the WV2+TSX combinations follow a same-side stereo configuration. For the center of the Munich study area, $\theta_{MEMPHIS} = 54^{\circ}$, $\alpha_{WV,1} = 12^{\circ}$, $\theta_{TSX} = 21^{\circ}$, $\alpha_{WV,2} = 8^{\circ}$. For the center of the Berlin area, $\alpha_{WV} = 10.3^{\circ}$, $\theta_{TSX} = 33^{\circ}$. Based on the analysis in Section \ref{sec:analysis}, the normalized height accuracies $\frac{\sigma_h}{\sigma_0}$ of these three datasets are 2.59, 1.04 and 1.30, respectively, which clearly shows the unfavorable intersection geometry experienced by a fusion of airborne SAR imagery (which typically shows a rather large off-nadir angle) and spaceborne optical imagery. For the purely spaceborne WV2+TSX case in Berlin, the intersection geometry is already better while that in Munich is almost optimal. Thus, although MEMPHIS provides by far the highest range accuracy $\sigma_R $ and the best matching success rate, the stereogrammetric fusion of this airborne SAR dataset with spaceborne WV2 imagery cannot provide as much an improvement in 3D reconstruction accuracacy as one would possibly expect.}

In consequence, this analysis shows the difficulty to trade-off a possible ease of matching and an effective stereo intersection geometry, as also indicated in the previous work of Toutin et al. \cite{toutin2000stereo}.

\subsection{Comparison of the Different Similarity Measures}
\begin{figure}[H]
\centering
\includegraphics[width= 0.8\columnwidth]{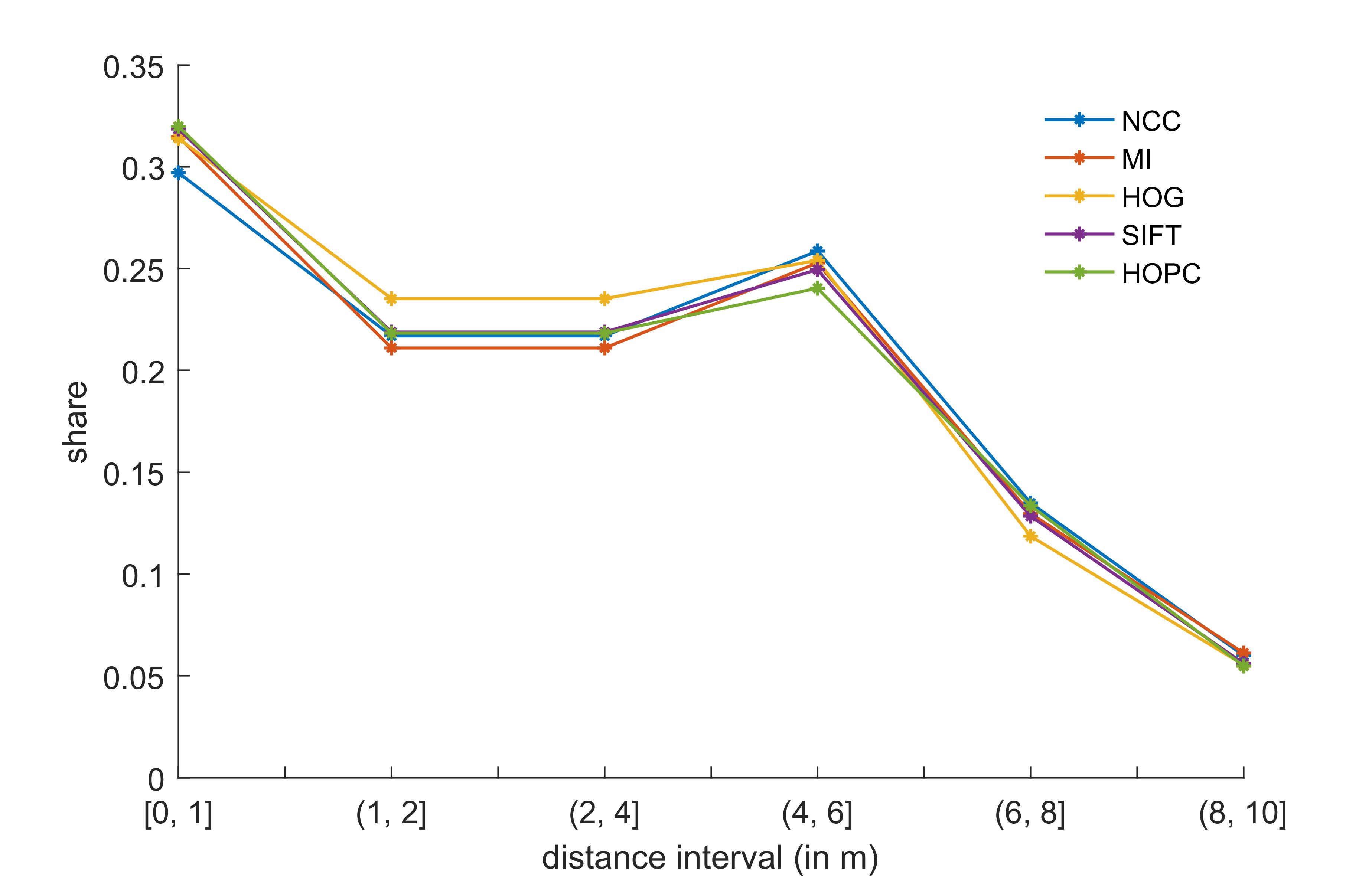}
\caption{The normalized histogram of the distance corresponding to different similarity measures. The distance are of all 1535 points within 10 meter from three experiments.}\label{fig:accuracyRatioSms}
\end{figure}

Comparing the results summarized in Tab.~\ref{tab:statisReconstruction} with respect to the different similarity measures, it becomes apparent that there exists no crucial difference between the individual measures, which confirms the impression gained from Fig.~\ref{fig:similarityMesuresEffect2Keypoints}: Whether a similarity measure works for a given key point depends on the nature of the % said 
key point and its image environment. \textcolor{black}{However, it has to be mentioned that the results in Tab.~\ref{tab:statisReconstruction} reflect what can be achieved after outlier removal by combination of all the individual similarity measures (cf. Section~\ref{sec:outlier}), so that the remaining difference between the individual similarity measures only reflects certain fine-positioning differences within the threshold window. Figure~\ref{fig:accuracyRatioSms} additionally compares the results before outlier removal. Here, it can be seen that the feature-based similarity measures that combine HOG-derivatives and an $L_2$-norm cost function slightly outperform the signal-based similarity measures.}

\textcolor{black}{Since none of the similarity measures employed in this work was specifically designed nor tuned for multi-sensor image matching tasks, this provides a hint towards critical future research directions: As first preliminary results confirm \cite{mou2017cnn}, future investigations should focus on the learning of a suitable similarity descriptor from exemplary data, as this approach has been shown to outperform hand-crafted approaches for purely optical image matching already \cite{Zagoruyko2015}.}

\subsection{Key Point Detection}%\label{sec:Discussion2}
\begin{figure*}[!tbh]
	\centering
    
	\subfigure[SAR image.]{
		%	\rule{4cm}{3cm}
		\includegraphics[height=5.8cm]{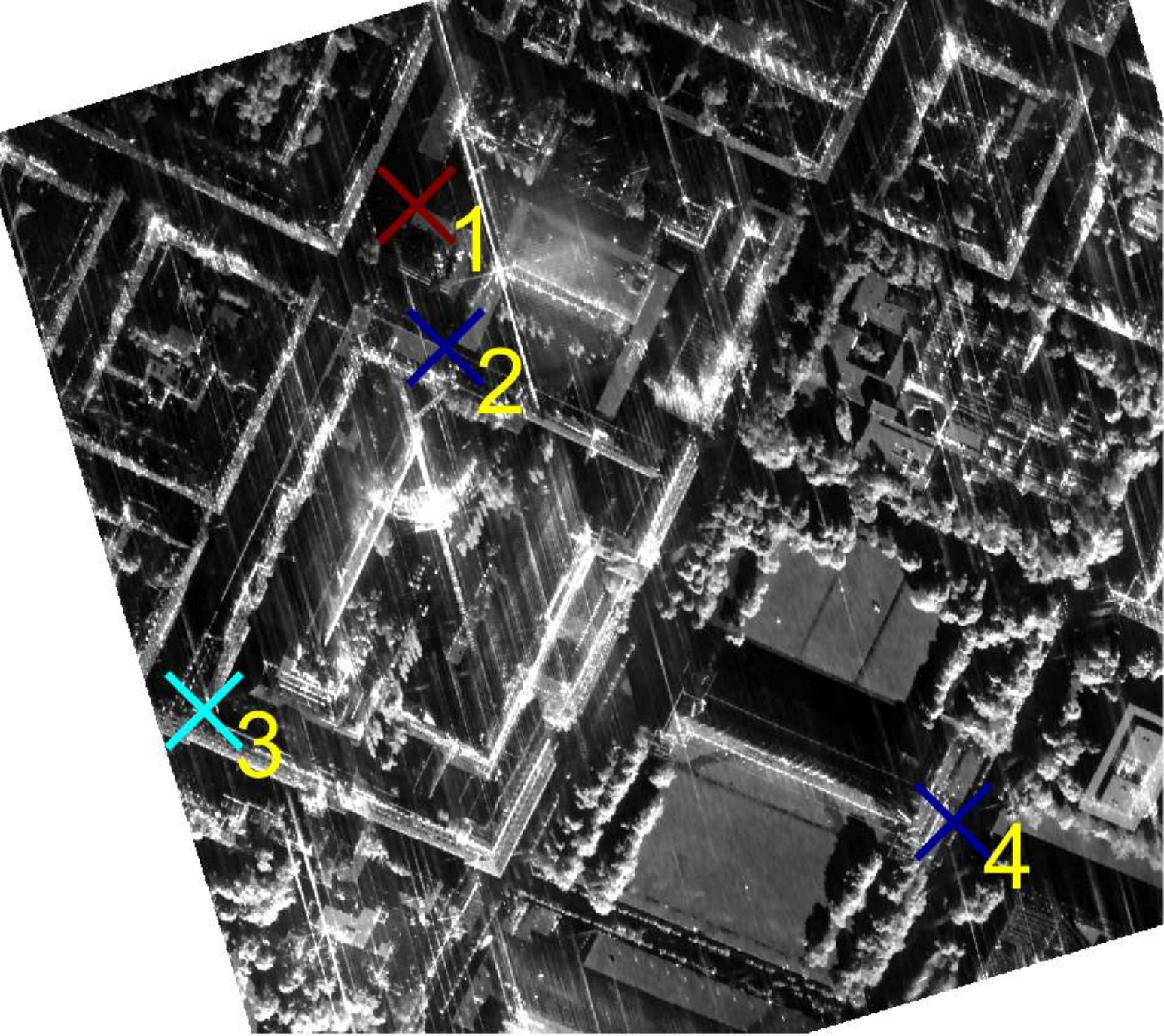}
		\label{}
	}%
	\subfigure[Optical image.]{
		%	\rule{4cm}{3cm}
		\includegraphics[height=5.8cm]{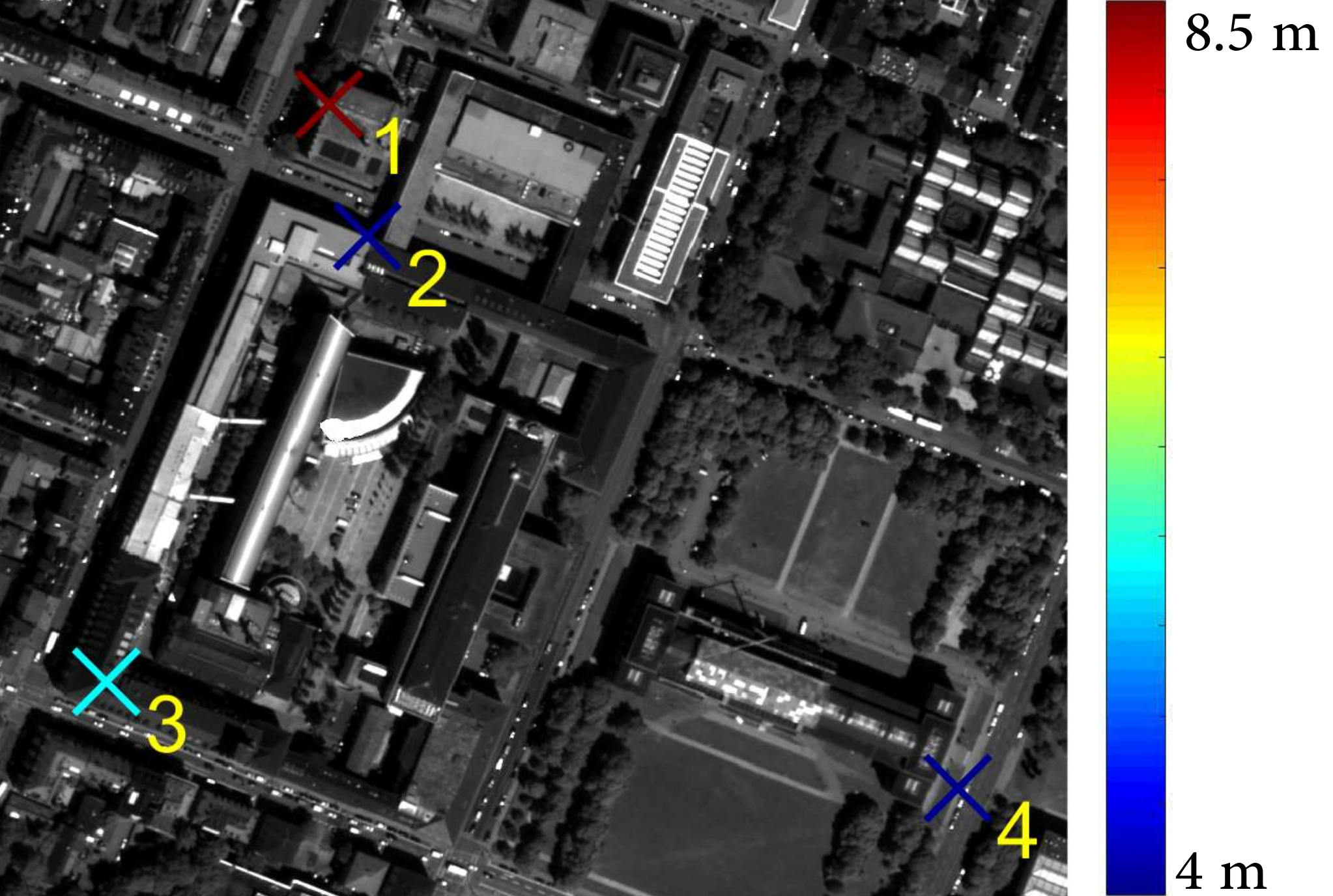}
		\label{}
	}
	\caption{The seven points with distances larger than four meters using NCC and data WV2+MEMPHIS.}
	\label{fig:tieLarge4m}
\end{figure*}

%\textit{Here, we can also discuss that a prior knowledge of the semantic contents can be beneficial by showing the distribution of reconstructed points with different accuracy}	
One of the core steps to influence the achievable stereogrammetry results for sparse matching situations is the key point detection step. As Fig.~\ref{fig:tieLarge4m} shows, \textcolor{black}{four out of 27} points matched in the WV2+MEMPHIS experiment are reconstructed with distances larger than four meters to the LiDAR reference data. All these points either lie %in areas containing trees appearing blurred in the despeckled SAR image, or 
in the surroundings of larger buildings, where mismatches can occur due to layover and shadowing. This indicates that there is a need for more sophisticated key point detection, which goes beyond a simple, signal-based corner detection and exploits, e.g. semantic knowledge about the objects present in the scene.

\section{Summary and Conclusion}
In this paper, we have investigated the potential and challenges present in the field of SAR-optical stereogrammetry when it comes to an analysis of complex urban areas with VHR imagery. We have theoretically derived optimal and unfavorable stereo geometries, which allows for an educated interpretation of experimental results. In addition, we have proposed a fully automatic procedure for a simultaneous solution to both the matching and the 3D-reconstruction problems, which exploits a search window constraint inspired by epipolar geometry. We have shown real data results based on \textcolor{black}{three} different experiments fusing spaceborne optical WorldView-2 data with both spaceborne and airborne SAR imagery acquired by TerraSAR-X and MEMPHIS, respectively. The results indicate that on the one hand image matching is less challenging for the high resolution airborne SAR data, while on the other hand spaceborne SAR data provides usually a more favorable intersection geometry, thus leading to comparable height reconstruction results.

While we were able to prove that fully automatic SAR-optical stereogrammetry for VHR imagery of urban scenes is generally possible with 3D-reconstruction accuracies in low meter-domain, we see the need for further research regarding two main issues: key point detection and measuring the similarity between SAR and optical image patches. If these two steps can be improved, more points with even better accuracies in the order of the pixel spacing will become feasible. In addition, potential errors in the sensor imaging parameters need to be considered in a multi-sensor bundle adjustment framework. In order to forgo the need for external ground control points (GCPs), the high geodetic accuracy provided by TerraSAR-X \cite{zhu2016geodetic} shall be exploited in that bundle adjustment framework to adjust the less-reliable optical orientation parameters.
Thus, besides general 3D reconstruction of urban areas, this research will also benefit applications such as GCP extraction from multi-sensor imagery.

 \section{Acknowledgements}
The authors want to thank everyone, who has provided data for this research: European Space Imaging for the WorldView-2 image, Fraunhofer FHR for the MEMPHIS image, Fraunhofer IOSB for the LiDAR reference data of Munich, and Land Berlin (EU EFRE project) and Landesamt für Vermessung und Geoinformation Bayern for the LiDAR reference data of Berlin.

This work is jointly supported by the China Scholarship Council, the Helmholtz Association under the framework of the Young Investigators Group SiPEO (VH-NG-1018, www.sipeo.bgu.tum.de), the German Research Foundation (DFG) under grant SCHM 3322/1-1, and the European Research Council (ERC) under the European Union's Horizon 2020 research and innovation programme (grant agreement No. ERC-2016-StG-714087, Acronym: \textit{So2Sat}).

 \bibliographystyle{elsarticle-num} 
 \bibliography{opticalSAR}

\end{document}